\documentclass[]{aa} 

\usepackage{graphicx}
\usepackage{txfonts}
\usepackage[utf8]{inputenc}
\usepackage{graphics}
\usepackage{amssymb}
\usepackage{lscape}
\usepackage{epsfig}
\usepackage{natbib}
\usepackage{longtable}
\usepackage{hyperref}
\usepackage[T1]{fontenc}
\usepackage[utf8]{inputenc}
\usepackage{multicol}
\usepackage[dvipsnames]{xcolor}
\usepackage{booktabs} 
\usepackage{csvsimple} 

\usepackage{lipsum}

\usepackage[normalem]{ulem}

\usepackage{caption}
\usepackage{subcaption}

\usepackage{booktabs}
\usepackage{hyperref}
\usepackage{multirow}
\usepackage{enumitem}

\usepackage[switch]{lineno}

\usepackage[dvipsnames]{xcolor}
\definecolor{myblue}{HTML}{1F77B4}
\definecolor{mygreen}{HTML}{2CA02C}
\definecolor{m}{HTML}{D62728}
\definecolor{mymagenta}{HTML}{D33682}
\definecolor{codepurple}{HTML}{C42043}

\hypersetup{
unicode=true,           
colorlinks=true,        
citecolor=myblue,       
urlcolor=black        
}
\begin{document}

   \title{XXL-HSC: Host properties of X-ray detected AGNs in XXL clusters}

 \author{E. Drigga\inst{1,2}
          \and
          E. Koulouridis\inst{1,3}
          \and 
          E. Pouliasis\inst{1}
          \and
          Y. Toba\inst{4,5,6}
          \and
          M. Akiyama\inst{7}
          \and A.~Ruiz\inst{1}
          \and
          C. Vignali\inst{8,9}
          \and
          I. Georgantopoulos\inst{1}
          \and
          T. Nagao\inst{6}
          \and
          S. Paltani\inst{10}
          \and M. Plionis\inst{11,2,12}
          \and M. Pierre\inst{3}
          \and
          B. Vijarnwannaluk\inst{5}
          }

   \institute{Institute for Astronomy \& Astrophysics, Space Applications \& Remote Sensing, National Observatory of Athens, GR-15236 Palaia Penteli, Greece\\
              \email{edrigga@noa.gr}
         \and
         Sector of Astrophysics, Astronomy \& Mechanics, Department of Physics, Aristotle University of Thessaloniki, Thessaloniki 54124, Greece
         \and
          Universit\'e Paris-Saclay, Universit\'e Paris Cit\'e, CEA, CNRS, AIM, 91191, Gif-sur-Yvette, France
         \and
         National Astronomical Observatory of Japan, 2-21-1 Osawa, Mitaka, Tokyo 181-8588, Japan
        \and
        Academia Sinica Institute of Astronomy and Astrophysics, 11F of Astronomy-Mathematics Building, AS/NTU, No.1, Section 4, Roosevelt Road, Taipei 10617, Taiwan
        \and
        Research centre for Space and Cosmic Evolution, Ehime University, 2-5 Bunkyo-cho, Matsuyama, Ehime 790-8577, Japan
        \and
        Astronomical Institute, Tohoku University, Aramaki, Aoba-ku, Sendai, Miyagi 980-8578, Japan
        \and
        Università di Bologna, Dip. di Fisica e Astronomia “A. Righi”, Via P. Gobetti 93/2, 40129 Bologna, Italy
        \and
        INAF – Osservatorio di Astrofisica e Scienza dello Spazio di Bologna, Via Gobetti 93/3, 40129 Bologna, Italy
        \and
        Department of Astronomy, University of Geneva, ch. d’Écogia 16, CH-1290 Versoix, Switzerland
        \and
        National Observatory of Athens, 18100 Thessio, Athens, Greece
        \and
        CERIDES, centre of Excellence in Risk \& Decision Sciences, European University of Cyprus, 2404 Egkomi, Cyprus
        }
\authorrunning{E. Drigga et al.}
\titlerunning{Properties of X-Ray AGNs in clusters}

   \date{\today}

  \abstract{ 
   There is compelling evidence that AGNs are strongly influenced by their environment, from their host galaxies to immense structures such as galaxy clusters. Therefore, studying the AGN population of clusters is essential, as both large-scale structures and AGNs play key roles in galaxy evolution, though the interactions between these elements are still not well understood.} 
   {The primary objective of this study is to unravel the different factors that may significantly affect the triggering of AGN activity in cluster galaxies, including galaxy merging and interactions with other galaxies, and ram pressure from the hot intracluster medium.}
   {For our purposes, we used 82 X-ray detected AGNs found within a $4r_{500}$ radius of 164 X-ray detected and spectroscopically confirmed galaxy clusters in the northern 25 deg$^2$ field of the XXL survey, up to a redshift of $z\sim$1. This field is also covered by deep optical observations of the Hyper Suprime-Cam, mounted on the 8m $\it Subaru$ Telescope, which allows for a reliable morphological classification of galaxies. We thoroughly investigated the morphology of X-ray AGN host galaxies, using both {\tt Statmorph} software and visual inspection, in an attempt to discover disturbances as indications of interactions that could lead to AGN triggering. Furthermore, using the X-ray hardness ratio, the optical spectra and the spectral energy distributions of the X-ray sources, we have studied the obscuration and other AGN properties, as well as the star formation rate of the hosts as further indicators of interactions.} 
   {We found a  moderately significant, at the $2\sigma$ confidence level, higher fraction of X-ray AGNs in galaxy clusters hosted by merging or disturbed galaxies, compared to non-active cluster galaxies or X-ray AGNs in the field. This excess is primarily localised in the cluster outskirts (between 1 and 2$r_{500}$).  
   Also, we discovered a higher number of  X-ray-hard (hence, possibly obscured) AGNs in clusters than in the field, at the $2\sigma$ confidence level, particularly in the outskirts. These findings further support the idea that galaxy mergers and interactions may serve as mechanisms for the triggering and obscuration of AGN activity.} 
    {The relatively high number of disturbed, merging, and possibly obscured AGN hosts in cluster outskirts suggests that galaxy merging and interactions are key drivers in triggering AGN activity in these outer regions of clusters.}

   \defcitealias{Adami2018}{XXL~Paper~XX}
   \defcitealias{Pierre16}{XXL~Paper~I}
   \defcitealias{Chiappetti18}{XXL~Paper~XXVII}
   \defcitealias{Lidman16}{XXL~Paper~XIV}
   \defcitealias{Koulouridis2018b}{XXL~Paper~XXXV}
   \defcitealias{Masoura2020}{XXL~Paper~XL}
   \defcitealias{Fotopoulou16}{XXL~Paper~VI}
   \defcitealias{Faccioli2018}{XXL~Paper~XXIV}
               
   \maketitle
%

\section{Introduction}

Supermassive black holes (SMBHs) are at the forefront of modern astrophysical research today not only because they are hosted by every massive galaxy in the local Universe, but also because the evolution of the SMBH and its host galaxy appears tightly linked \citep[e.g.][]{Gultekin09,Zubovas2012}. All SMBHs are thought to undergo active phases, the so-called active galactic nucleus (AGN) phases, during which they accrete the surrounding gas and emit an immense amount of energy. Theoretical models and simulations have proposed that during this active phase, the active nucleus produces a feedback wind that can explain the co-evolution of the SMBH and its host galaxy  \citep[e.g.][]{Schawinski2009,Cen11}. Therefore, the study of AGNs is essential for understanding the cosmic history of accretion into SMBH and their relation to the host galaxy.

One specific research direction for investigating the cosmic history of AGN evolution is studying AGNs as a function of their environment. Several studies have provided evidence that AGNs are affected by both their immediate surroundings \citep[e.g.][]{Maiolino1997,dultzin99,Sorrentino2006A,Gonzales2008,Silverman08,Dultzin2008,Koulouridis2006b,Koulouridis2006a,Koulouridis2013,Manzer2014,Silva2021,Duplancic2021,Pierce2023,Li2023} and by their large-scale environment \citep[e.g.][]{Constantin2008,Stroe2020,Ceccarelli2022, Hashiguchi, Munoz2023,Koulouridis16b,Koulouridis2024,Toba24,deVos2024}. In addition, early studies reported overdensities of X-ray point-sources in clusters with respect to the field \citep{Cappi01,Molnar02,DElia04,branchesi2007,Gilmour2009}, and others have spectroscopically verified the existence of a large population of AGNs in clusters and argued on their probable evolution with redshift \citep[e.g.][]{Martini02,Johnson2003,Martini07,Martini09}. Therefore, it is crucial to thoroughly investigate the AGN population of galaxy clusters, as both the immense structure and the powerful nucleus seem to play an important role in galaxy evolution.

However, the interplay among the immense cluster and the powerful nucleus is still not well understood. This uncertainty arises from the various physical mechanisms that may influence galaxies and SMBHs within clusters. Numerous studies have demonstrated that the AGN fraction in member galaxies of massive clusters ($M>10^{14}M_\sun$) is lower than the respective fraction in field galaxies \citep[e.g.][]{Kauffmann2004,Gavazzi2011,Ehlert2013,Ehlert2014,Mishra2020,Beyoro2021}. This is likely caused by ram pressure stripping (RPS). In more detail, galaxies within the dense intracluster medium (ICM) of clusters are subjected to intense  pressure, leading to the efficient ram pressure stripping of their gas. This reduces the availability of cold gas necessary for fueling AGNs, resulting in a diminished cold gas reservoir required to trigger nuclear activity \citep[e.g.][]{Gunn72,Cowie77,Giovanelli85,Popesso06b,Chung2009,Haines12,Sabater2013,Jaffe15,Poggianti2017b}. The impact of RPS is expected to be proportional to the cluster's mass and inversely proportional to the galaxy's mass \citep[e.g.][]{Boselli2022}. In support of these expectations, studies on poor clusters and groups reported that AGN activity in group galaxies is at least as frequent as in the field \citep[hereafter \citetalias{Koulouridis2018b}]{Sabater2012, Koulouridis14, Koulouridis2018b}. However, \citet{Bufanda2017} did not detect differences in the fraction of X-ray luminous AGNs ($L_{\rm X}>10^{43}$ erg sec$^{-1}$) between groups and clusters for 432 clusters from the Dark Energy Survey (DES) up to $z=0.95$. We note that \citet{Poggianti2017b} proposed that RPS might also act as a triggering mechanism for AGN activity in cluster members. Furthermore, the so-called jellyfish galaxies \citep{Chung2009,Bekki2009,Poggianti2017b}, which are conspicuously affected by RPS, were found to host a significantly higher number of AGNs than similar field galaxies \citep{Peluso2022}. Finally, many studies have revealed a positive evolution of the AGN fraction in cluster galaxies with redshift \citep[e.g.][]{Kocevski2009,Fassbender2012,Martini2013,Bufanda2017,Hashiguchi}, while low-mass protoclusters at higher redshifts may potentially contain a higher number of AGNs \citep[][]{Lehmer13,Krishnan17,Gatica2024, vito24}.

Contrary to the suppression of AGNs observed in the centre of massive clusters, an increase has been found in the outskirts of the cluster \citep[e.g.][]{Johnson2003,branchesi2007,Koulouridis14}. However, the results vary depending on the different selection of clusters and AGN samples. In particular, \citet{Ruderman2005} discovered a mild excess of X-ray sources between 1.5 and 3 Mpc in massive clusters that span the redshift range of z$ = 0.3 - 0.7$. However, the excess was found only in dynamically relaxed clusters, while no excess was found in the outskirts of disturbed clusters. These findings were confirmed more recently in the optical band by \citet{Stroe2021}, in a sample of 14 clusters ($z\sim 0.15-0.31$) that span a wide range of masses and dynamical states. They found that the H$\alpha$-detected AGN fraction peaks in the outskirts of relaxed clusters ($\sim 1.5-3$ Mpc). In addition, \citetalias{Koulouridis2018b} revealed a significant overdensity of spectroscopically confirmed X-ray AGNs in the outskirts of low-mass clusters ($M_{500}$\footnote{The $M_{500}$ cluster mass refers to the total mass of a galaxy cluster within a spherical region where the average density is 500 times the critical density of the universe at the cluster’s redshift. The corresponding radius, $r_{500}$, is the radius at which this density contrast is reached.} $<10^{14} \, M_\sun$ and $0.1<z< 0.5$) from the XXL Survey \citep[hereafter \citetalias{Adami2018}]{Adami2018}, while no excess was confirmed for higher cluster masses. At higher redshifts, a similar excess of X-ray AGNs was also reported by \citet{Fassbender2012} between 4 and 6 arcmin from the centres of 22 massive clusters ($0.9 < z < 1.6$). \citet{Koulouridis2019} confirmed a highly significant excess, at the 99.9\% confidence level, of X-ray point-like sources in the outskirts ($2-2.5r_{500}$) of the five most massive, $M_{500}^{SZ}>10^{14} M_{\odot}$ , and distant, z$\sim$1, galaxy clusters in the \textit{Planck} and South Pole Telescope (SPT)\textit{} surveys. Very recently, \citet{Koulouridis2024} reported a significant excess of X-ray AGNs in the outskirts of relaxed clusters, compared both to non-relaxed clusters and to the field. Furthermore, a similar excess in cluster outskirts ($\sim3r_{500}$) was also recently found in the {\it MAGNeticum} simulations \citep{Rihtarsic2023}. Nevertheless, some studies have  found that AGNs have no special position inside galaxy clusters \citep[e.g.]{Gilmour2009, Ehlert15}, unless only the most powerful optical AGNs are considered. In addition, \citet{Munoz2023} demonstrated that the observed excess of X-ray sources in the outskirts of massive clusters ($M > 5\times10^{14} \, M_\sun$) at $z\sim1$ might be caused by projection effects.

The increase reported by various works may be attributed to a corresponding increase in the galaxy merging rate, which is favoured by the lower galaxy velocities in the outskirts compared to the centre, as well as low-mass groups ($M_{500} < 10^{14},  M_\sun$) when compared to massive clusters \citep[e.g.][]{Ehlert15, Lopes2017, Gordon2018}. During a merger, the involved galaxies experience tidal forces that can lead to the redistribution of their stellar and gaseous components, triggering star formation and potentially fueling AGNs. Generally, merging represents a significant process in the hierarchical formation and evolution of galaxies. As galaxies traverse the dense environments of clusters, gravitational interactions and dynamical friction facilitate their coalescence. In addition, notably in the outskirts of clusters, galaxies experience less intense ICM pressure, preserving their cold gas supplies and thus maintaining the conditions favourable for AGN activity. Furthermore, nuclear activity in the outskirts may be triggered by interactions between the host galaxy and the cluster itself during passage through virial shocks \citep[e.g.][]{Keshet18}.

Another possibility, which explains the excess of AGN in the outskirts of clusters, is that AGNs enter the cluster environment along with infalling small groups. Galaxies typically do not enter clusters as isolated entities. Instead, they often enter as part of smaller groups that offer a more favourable environment for AGN triggering, namely, they are pre-processed. Preprocessed indicates that these galaxies have already experienced significant interactions and evolution before entering the cluster \citep[e.g.][]{Fujita2004,Haines2015,Sengupta2022,Lokas2023}. However, \citet{Koulouridis2024} found no evidence that the the prevalence of X-ray AGNs in clusters is affected by the presence of X-ray-detected infalling groups, or that their location is correlated with the positions of these infalling groups. 

In summary, the environmental dichotomy between the cluster centre and the outskirts results in a spatial variation of AGN activity. However, the physical mechanisms that trigger nuclear activity in cluster galaxies are still debated, since the AGN frequency appears to be affected by multiple factors. In this context, we have studied the morphology of cluster galaxies hosting X-ray-detected AGNs up to a redshift of $z\sim1$. They are located out to a distance of $4r_{500}$ radii from the centres of X-ray detected clusters in the northern 25 deg$^2$ XXL survey field, which is also covered by deep Hyper Suprime-Cam (HSC) optical observations. Our aim is to investigate the role of galaxy merging and interactions in the triggering of AGN activity. In addition, we have studied the obscuration and the accretion power of the AGNs, as well as the star formation rate (SFR) of the hosts, as further indicators of the physical mechanisms that drive their triggering and evolution.

The outline of this paper is as follows. In Sect. \ref{sec:sample} we discuss the data preparation and sample selection. The methodology is described in \ref{sec:method}, and the results are presented in Sect. \ref{sec:results}.  Our discussion and conclusions are presented in Sect. \ref{sec:disc}, and a summary is provided in Sect. \ref{sec:conc}. Throughout this paper we assume a Planck cosmology \citet{Planck2016} of H$_{o}=67.8$ h km s$^{-1}$ Mpc$^{-1}$, $\Omega_{m}=0.308$, and $\Omega_{\Lambda}=0.692$.

\section{Sample selection}
\label{sec:sample}

In this section, we describe the surveys and the data used to study the properties of X-ray AGNs in galaxy clusters of the XMM-XXL northern field. We used the latest X-ray catalogue from the XXL survey along with deep HSC imaging. Additionally, we exploited all available spectroscopic and photometric data from a multitude of surveys, to study the spectral properties and the spectral energy distributions (SEDs) of our sources.

\subsection{XMM-XXL survey}

The XXL survey \citep{Pierre16} is the largest {\it XMM-Newton} programme to date totaling $\sim6.9$ Msec. It covers two extragalactic fields of about 25 deg$^2$ each, XXL-North (XXL-N) and XXL-South (XXL-S), at a point-source sensitivity of $6\times10^{-15}$ ergs$^{-1}$ cm$^{-2}$ in the [0.5-2] keV band (completeness limit). The main goals of the survey are to provide constraints on the dark energy equation of state from the space-time distribution of clusters of galaxies and to serve as a pathﬁnder for future, wide-area X-ray missions. In the current work, we have used the latest XXL v4.3 catalogue. The cluster selection criteria are described in \citep{Pacaud06} and the 365 cluster catalogue is presented in \citetalias{Adami2018}. The creation of the XMM-XXL X-ray point-source catalogue, along with the optical counterpart matching and associated multi-wavelength data, is described in \citet{Chiappetti18}, also known as \citetalias{Chiappetti18}. However, in this work we used the more recent internal release of the catalogue obtained with the V4.2 XXL pipeline that contains in total 15547 X-ray sources. 

Spectroscopic redshifts in XXL-N were obtained with large spectroscopic surveys, such as SDSS, VIPERS \citep{Guzzo14}, and GAMA \citep{Liske15}, and from a large campaign with the AAOmega spectrograph mounted on the Anglo-Australian Telescope described in \citet{Lidman16}, also known as \citetalias{Lidman16}. Other smaller-scale spectroscopic observations \citep[e.g.][]{Koulouridis16b} complement the sample. The photometric redshifts used are described in \citet{Fotopoulou16}, also known as \citetalias{Fotopoulou16}.

\subsection{Hyper Suprime-Cam (HSC)}
The HSC \citep{miyazaki18} is a 1.77 deg$^2$ imaging camera with a pixel scale of 0.168 arcsec mounted at the prime focus of the 8.2m {\it Subaru} Telescope. It is composed of 116 charge-coupled devices (CCDs; 104 for science, four for the auto guider, and eight for focus monitoring). This facility is operated by the National Astronomical Observatory of Japan on the summit of Maunakea (Hawaii, USA). The HSC Subaru Strategic Program (HSC-SSP; \citep{aihara18}) is a three-layered survey (wide, deep, and ultradeep) of 1400 deg$^2$ in five different bands (grizy) and four narrow filters. For our purposes, we used imaging and photometric data from the deep HSC survey, which covers most of the northern XMM-XXL survey, reaching a depth of $r\sim27$.

\begin{figure*}[htp]
    \centering
    \includegraphics[width=18cm]{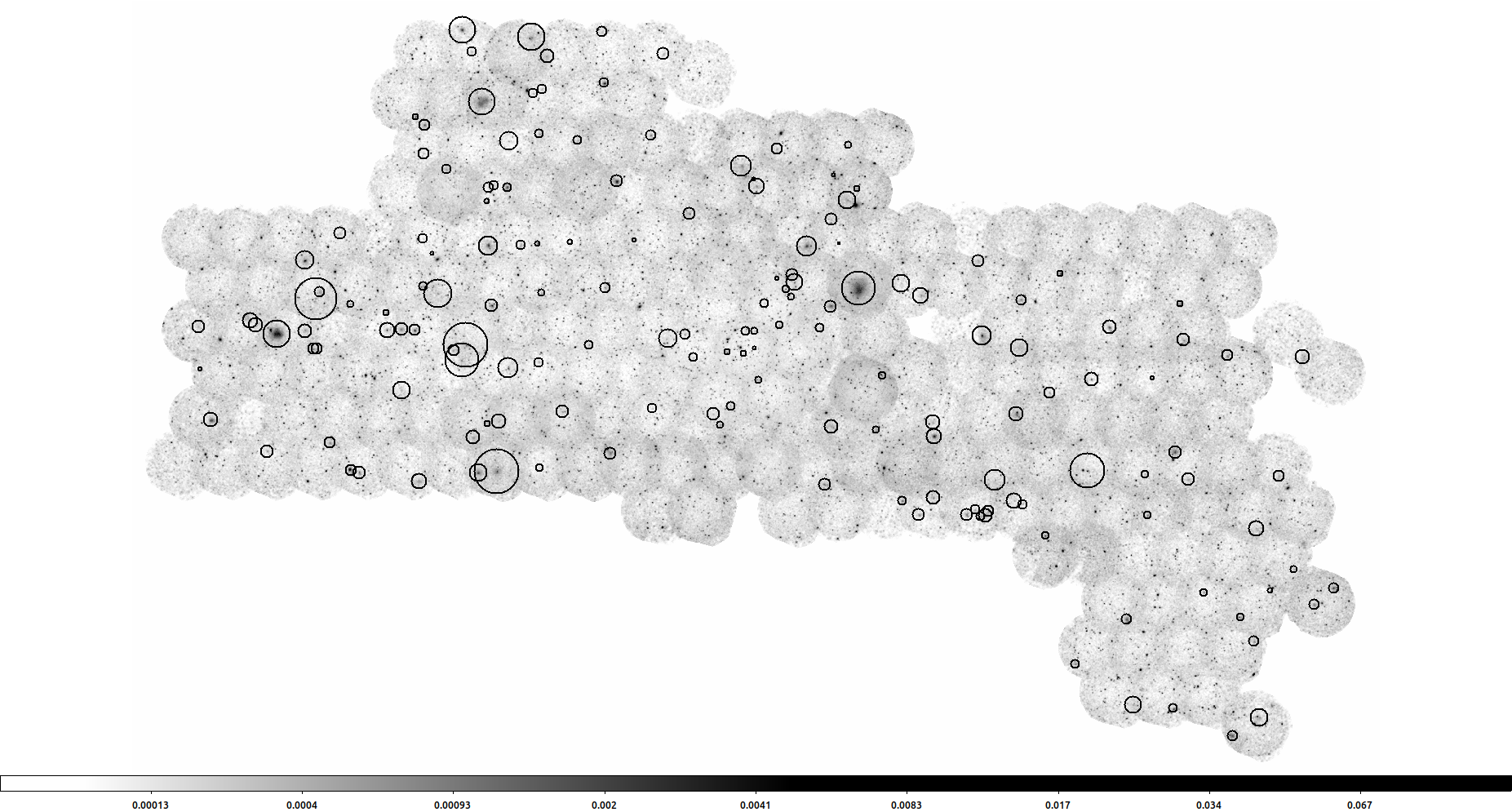}
    \caption{XXL-N count-rate map. We overlay the positions of 164 X-ray detected clusters used in this paper. Circles mark the position and they represent the $r_{500}$ radius of each cluster.}
    \label{fig:xxl}
 \end{figure*}   

 \subsection{X-ray detected AGNs in XXL clusters}
 \label{sub:selection}
 
 Our aim was to combine X-ray data from the XXL survey with HSC imaging to perform a reliable visual inspection of the X-ray AGN host galaxies up to a high redshift ($z\sim1$). The HSC footprint covers only the northern XXL field, therefore our sample comprises 164 clusters. The frequency of AGN activity in most of these clusters, up to redshift $z=0.5$, was studied thoroughly in \citetalias{Koulouridis2018b}. We initially selected all point-like X-ray sources within a projected cluster-centric distance of 4$r_{500}$. We used the $r_{500,MT}$ values when available, from Table 5 in \citetalias{Adami2018}, derived after a spectral fit of the X-ray observations. In the case where a spectral fit was not possible, we used the $r_{500,scal}$ values derived from scaling relations \citepalias[][Table F.1]{Adami2018}. The XXL-N field and the location of the 164 X-ray detected clusters are presented in Fig.~\ref{fig:xxl}.
 
\begin{figure}[htp]
    \centering
    \includegraphics[width=8cm]{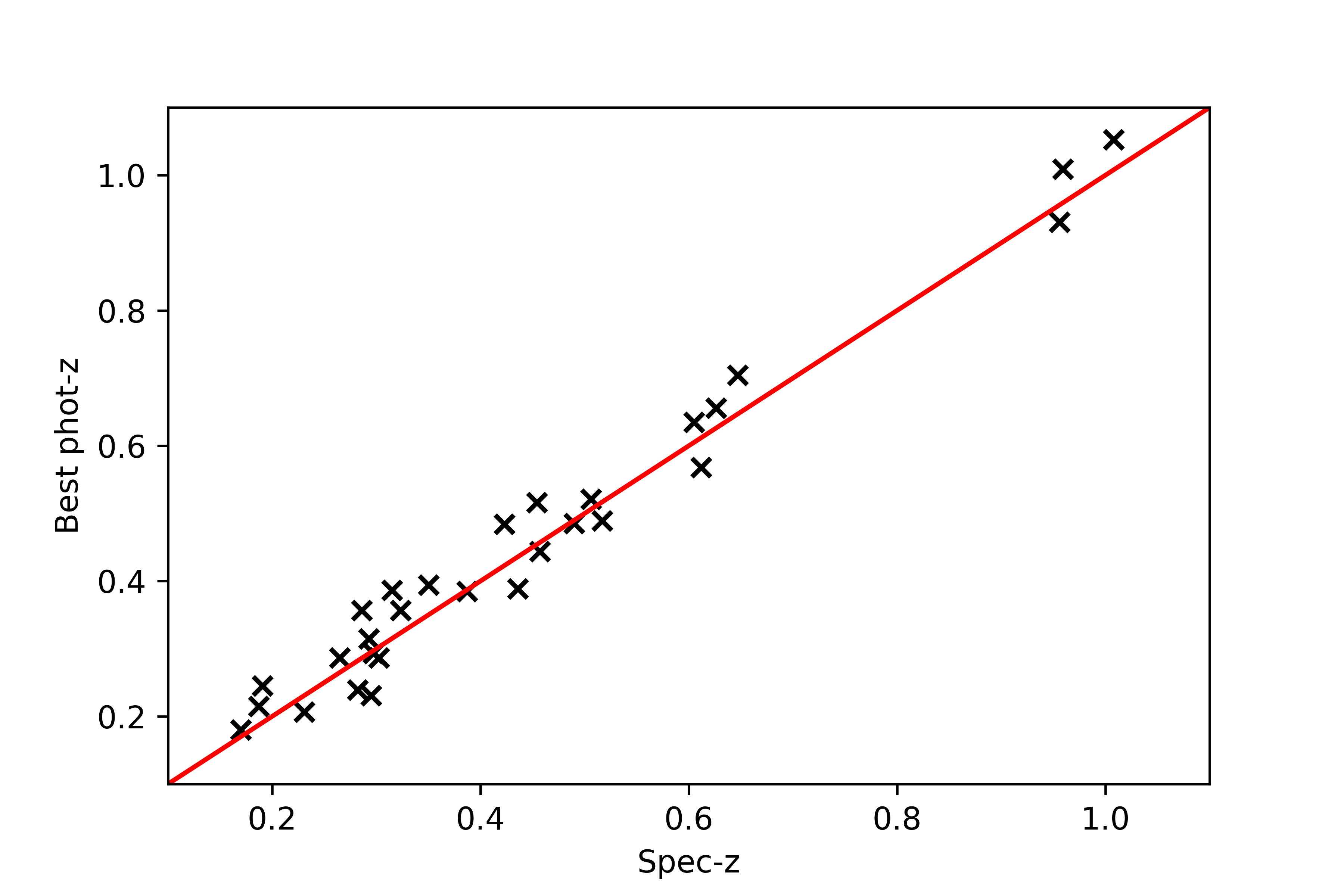}
    \caption{Spectroscopic versus photometric redshift scatter plot of the spectroscopically confirmed X-ray detected AGNs in clusters. The sources were selected based on specific photometric redshift criteria, as described in Sect.~\ref{sub:selection}. The red line denotes the equality. }
    \label{fig:specphottype}
\end{figure}

 Subsequently, only sources above a luminosity threshold of $L_{X[0.5-10]\,\rm{keV}} > 10^{42}$ erg s$^{-1}$ at the redshift of each cluster were kept in the sample. This luminosity threshold indicates that X-ray emission from point-like sources is most likely due to an AGN rather than other sources such as X-ray binaries or star formation. Finally, only galaxies with concordant spectroscopic or photometric redshift with the redshift of the clusters were kept in the sample. In more detail, in case of available spectroscopic redshifts ($\sim$50\% of our sample), we require the line-of-sight (los) relative velocity to the average cluster redshift, $\Delta\upsilon=\upsilon_{los}-\langle\upsilon\rangle$, to be less than 2000 km/sec. When only a photometric redshift is available, the relative los distance should not exceed $0.1(1+z_{cl})$, where $z_{cl}$ is the average cluster redshift. This threshold allows the selection of the sources with the most reliable photometric redshifts \citepalias[][see figure 2]{Fotopoulou16}, while it excludes catastrophic outliers. Furthermore, if the source was outside the 68\% confidence interval of the photometric redshift probability distribution (PDZ) around the median value, or if this interval was larger than 0.5, the source was excluded from our analysis.  The latter criterion ensures that sources with flat PDZs, and therefore unreliable redshift estimation, are excluded from the analysis. In Fig.~\ref{fig:specphottype} we plot the spectroscopic vs. photometric redshifts for the sources that would have been included in our sample even if the spectroscopic redshift was not available. While our stringent criteria verify the purity of our sample, some true X-ray AGNs that belong to the clusters inevitably will be missed.  

\begin{figure}[htp]
    \centering
    \includegraphics[width=10cm]{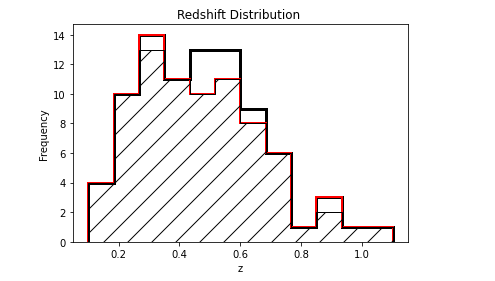}
    \caption{Normalised redshift distribution. The hatched area corresponds to the main sample of 82 X-ray detected AGNs in clusters while the black line and the red line correspond to the 166 field X-ray AGNs (sub-sample ii) and the 208 clusters galaxies (sub-sample iv), respectively.}
    \label{fig:stellarmass}
 \end{figure}

 The final selected main sample comprises 82 X-ray AGNs up to 4$r_{500}$ radii from the centres of the clusters. Moreover, we assembled four control samples to enable a thorough comparative analysis: (i) 1987 field X-ray-detected AGNs in XXL-N and (ii) 1914 cluster galaxies selected based on their spectroscopic redshift within 4$r_{500}$ radii of our 164 clusters. From these two control samples, we extracted two smaller sub-samples of (iii) 166 field X-ray AGNs and (iv) 208 cluster galaxies respectively, in order to visually inspect their morphology and optical spectra. Control samples (i) and (iii) for field X-ray AGNs are formulated to mirror the main sample in terms of both redshift and X-ray luminosity distribution. Control samples (ii) and (vi) for cluster galaxies are designed to emulate the main sample in terms of redshift and stellar mass distribution. A summary of the samples is presented in Table~\ref{table:samples}, while the redshift distributions of the main sample and the two small subsamples (iii) and (iv) are presented in Fig.~\ref{fig:stellarmass}.

\begin{table}
\caption{Summary of the samples}             
\label{table:samples}      
\centering                          
\begin{tabular}{l c c}     
\hline\hline                 
Sample & description & size \\    
(1) & (2) & (3) \\
\hline                        
   Main & cluster X-ray AGNs & 82\\      
     control (i)&  field X-ray AGNs & 1987 \\ 
    control (ii) & cluster galaxies & 1914   \\ 
    control (iii) & field X-ray AGNs & 166  \\ 
     control (iv) & cluster galaxies & 208  \\            
\hline                                   
\end{tabular}
\tablefoot{Control samples (iii) and (iv) were directly derived from the larger respective samples (i) and (ii) in a manner that preserves their properties and enables visual inspection.}
 \end{table}

\section{Methodology}
\label{sec:method}

In this section we describe the methodology we followed to treat our data. In order to examine in detail the properties of X-ray AGNs in clusters, we divided the circum-cluster area into four concentric annuli centred on the X-ray peak of the diffuse emission. We selected each annulus to have a width of $r_{500}$ radius. Employing any $r_\Delta$ radius is crucial when studying the impact of the cluster environment on AGNs and their host galaxies, as it provides a direct link to the physical conditions at each galaxy's location \citep{Koulouridis2019}. We can assume that similar conditions prevail within the specific annulus of other clusters, regardless of the actual physical or projected distance. The $r_{500}$ radius is also useful for direct comparison with previous results as it is used extensively in the literature. Then, we stacked the number counts of X-ray AGNs found in each respective annuli of all clusters.  We consider the first $r_{500}$ annulus to be the centre of the cluster, while the second annulus the cluster outskirts. The motivation to use these specific boundaries comes from the fact that the radius of 2$r_{500}$ roughly coincides with the virial radius and the splash-back radius\footnote{The splash-back radius of a galaxy cluster is the boundary that marks the outer edge of its gravitational influence, where the accreted matter reaches its farthest point after falling into the cluster for the first time.}. Therefore, we can assume that galaxies within this radius are bound to the cluster potential, while outside they are not yet influenced by any effect of the dense cluster environment. Nevertheless, for any comparison with the field in the current work we have used independent control samples of X-ray AGNs (see Sect.~\ref{sub:selection}). We note that within an area roughly encompassing the inner half of the $r_{500}$ radius, therefore $\sim$25\% of the total area of the central annulus, the high X-ray background caused by the diffuse ICM emission may hinder the detection of low-luminosity point-like sources \citep[e.g.,][]{Bhargava2023}.

\subsection{Morphological analysis of X-ray AGN host galaxies}
\label{sub:morph}

\begin{figure}[htp]
    \centering
    \includegraphics[width=9cm]{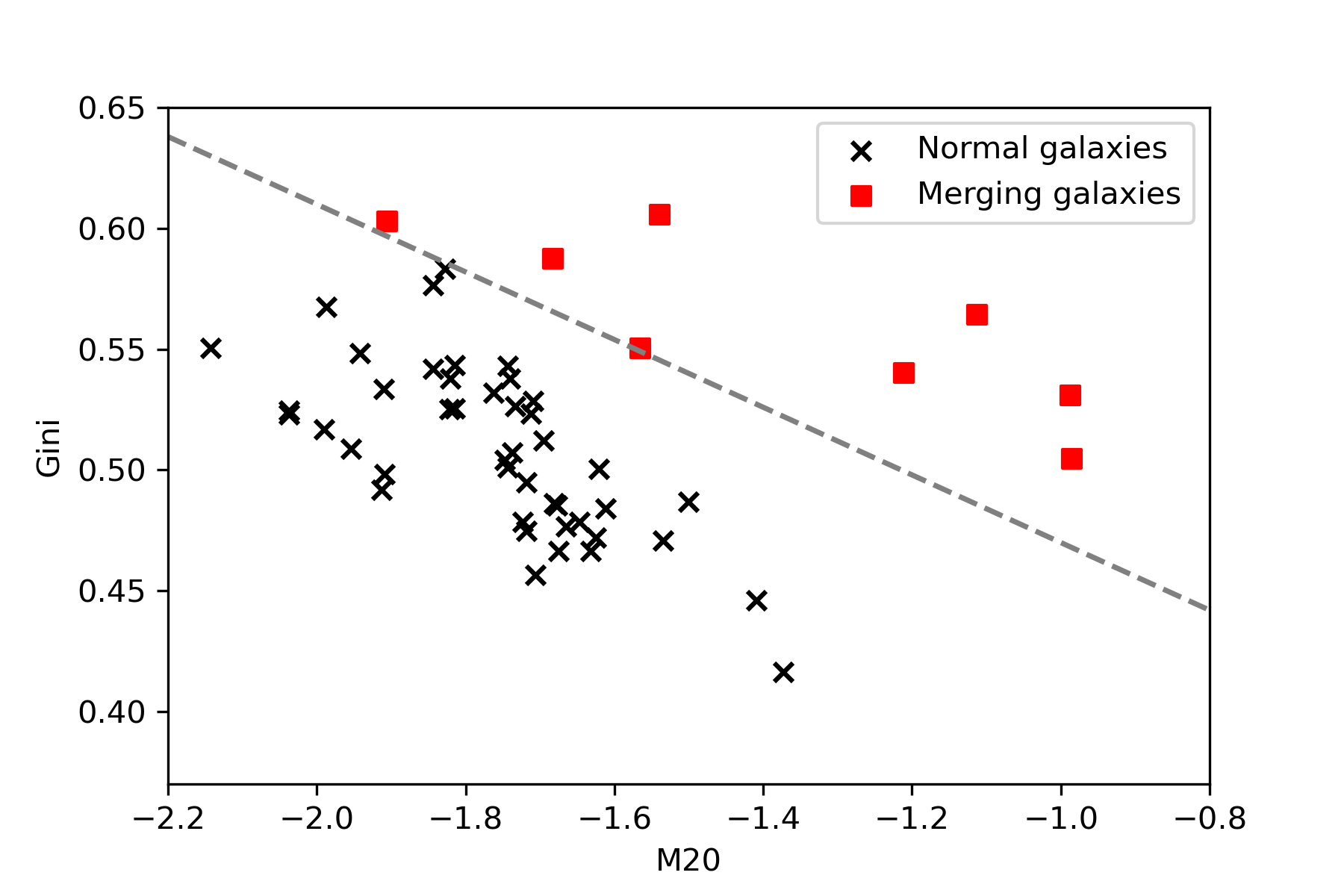}
    \caption{Gini-M20 classification diagram of galaxy morphology. Black crosses and red squares mark the position of the X-ray AGNs in clusters; we excluded those with bad measurements. The hatched line (equation \ref{eq:1}) separates "normal" and merging galaxies.}
    \label{fig:giniM20}
\end{figure}

Our goal is to investigate whether merging is more frequent in X-ray AGN hosts within clusters than in the field or when compared to inactive cluster galaxies. To this end, we used HSC imaging and photometry in order to examine and determine the morphology and the immediate environment of all galaxies in our samples. Speciﬁcally, we were investigating for signatures of mergers, or interactions with neighbouring galaxies. 
To find merging systems in all samples, we used {\tt Statmorph} \citep{Rodriguez}.
This code is designed for measuring the morphological properties of galaxies, especially for the non-parametric morphological indicators.
In this work, we employed Gini coefficient and $M_{20}$ diagnostics to select mergers \citep{Lotz}.
The Gini coefficient evaluates the bias of the light distribution in a galaxy, where larger values indicate a more concentrated light distribution. On the other hand, $M_{20}$ assesses the distribution of the brightest twenty percent of the galaxy's light, where higher values suggest more extended bright regions.
To find the merging systems, we applied the Gini-$M_{20}$ classification to the HSC $i$-band image (see Yanagawa et al. in prep. for a full description of merger classification for HSC sources with {\tt Statmorph}). In particular, we classified as mergers all galaxies that fall above the line that divides merging and non-merging systems \citep{Lotz2008, Rodriguez}: 
\begin{equation}
\label{eq:1}
Gini = -0.14 M_{20} + 0.33
\end{equation}
Sources that were flagged as having a problem with basic measurements for various reasons  \citep[][see section 4]{Rodriguez} were discarded from the above analysis as undefined (U). In Fig.~\ref{fig:giniM20} we plot the positions of the X-ray AGN host galaxies on the Gini-$M{20}$ diagram and the dividing line.

\begin{figure}[htp]
    \centering
    \includegraphics[width=7.5cm]{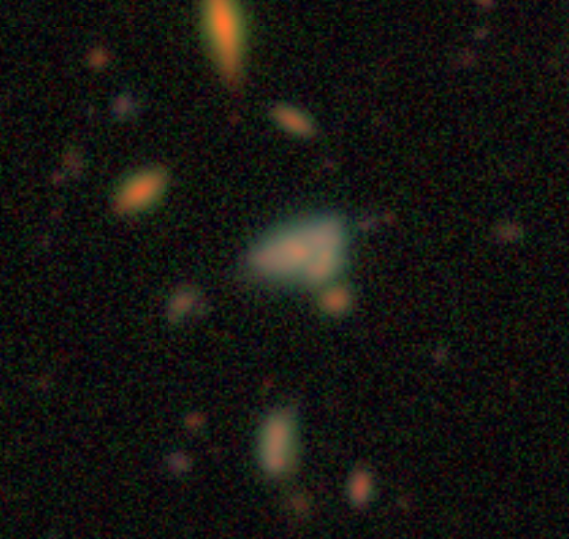}
    \includegraphics[width=7.5cm]{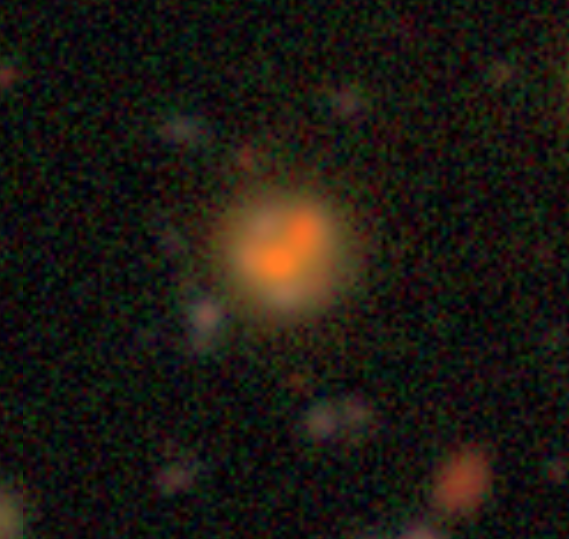}
    \includegraphics[width=7.5cm]{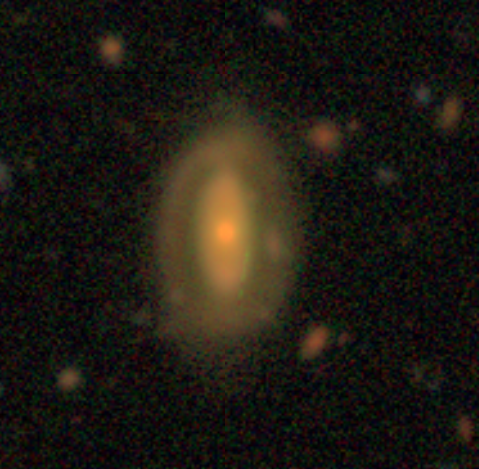}
    \caption{RGB colour images based on the \textit{gri} filters of HSC. Top panel: X-ray AGN merging host of 4XLSSU J023147.2-045702 with a disturbed morphology (z=0.190). Middle panel: X-ray AGN merging host of XLSSU J023324.9-053106 with two distinct cores (z=0.436). Bottom panel: X-ray AGN disturbed host of XLSSU J021649.4-032831 (z=0.21). The image scale is approximately 30"$\times$30".}
    \label{fig:mergers}
\end{figure}

Additionally, for the main sample of 82 X-ray AGNs and for the two small control subsamples (iii) and (iv) we performed a visual examination of all HSC images and classified the AGN host galaxies as merging or disturbed (M), Ellipticals (E), Spirals (S) and Undefined (U). In more detail, we classified galaxies as disturbed (D) if they showed clear signs of disturbed morphology, the presence of tidal tails, or noticeable substructures. In addition, mergers should exhibit two visibly separate cores. We note that a few galaxies classified as "U" by {\tt Statmorph} due to poorly masked secondary sources were recovered as merging systems when the redshifts of the two cores were also taken into account. Conversely, cases of projection that were incorrectly identified as mergers were excluded. Thus, while the results of {\tt Statmorph} and visual classification are not identical, they give consistent statistical results. Visually small galaxies or bright QSOs are classified as undefined. However, only two cases among the 82 X-ray AGNs in the main sample exhibit such powerful QSO activity that it obstructs the morphological classification of their host galaxies. Three examples of X-ray AGN merging or disturbed hosts can be found in Fig.~\ref{fig:mergers}. A mosaic of the full sample of the 82 X-ray-detected AGNs in clusters can be found in Table~\ref{fig:mosaic1}.

\subsection{Multi-wavelength properties}

\subsubsection{Optical spectra analysis}
\label{sub:optical}

Active Galactic Nuclei can be classified into two main types, broad-line (BL) and narrow-line (NL) AGNs, according to the width of Balmer emission lines in their optical spectrum. In more detail, in AGN models a torus of gas and dust surrounds the central SMBH, obscuring the nucleus from certain viewing angles. Narrow-line AGNs are believed to be obscured by the torus, which blocks the direct radiation from the central AGN region so that the broad components of the Balmer lines cannot be detected. However, many studies suggest that below a specific accretion rate of material into the black hole the broad-line region (BLR) might be absent \citep{Nicastro2000,Nicastro2003, Bian2007,ElitzurHo2009,Bianchi2012,Bianchi2017,Elitzur2014,elitzur2016,Koulouridis2014, Koulouridis16a}. Therefore, some of the low-luminosity sources of our sample might be "true" narrow-line AGNs, intrisically lacking the BLR. In particular, almost 25\% of our sources satisfy the criteria of accretion rate and luminosity described in \citet{Marinucci2012,Koulouridis16a} that would classify them as potentially lacking the BLR. Furthermore, additional obscuration or dilution, caused by the host galaxy, might affect the AGN classification, regardless of the inclination of the torus \citep[e.g.][]{Lagos2011,Gkini2021}  

\begin{figure}[htp]
    \centering
    \includegraphics[width=9cm]{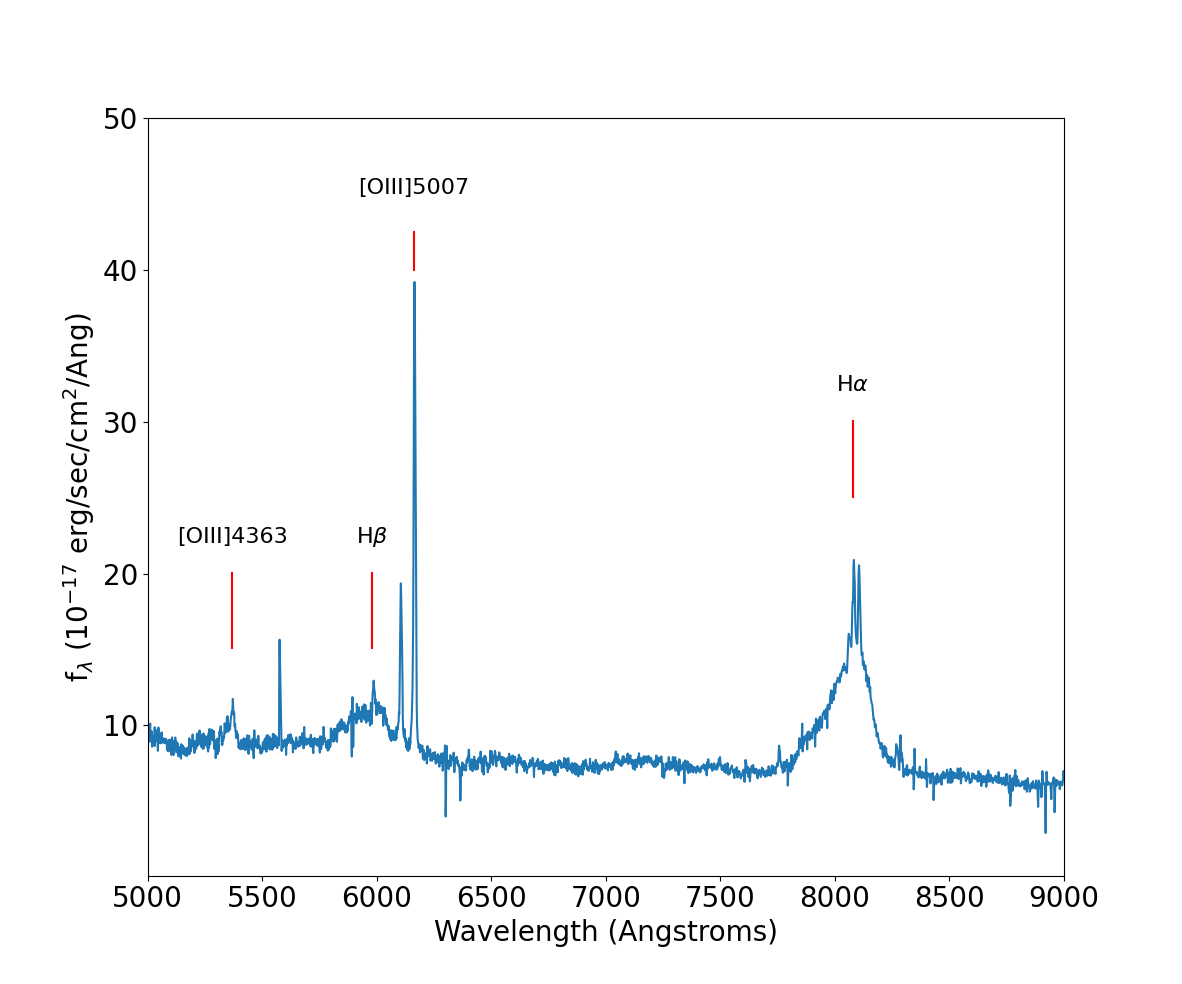}
    \includegraphics[width=9cm]{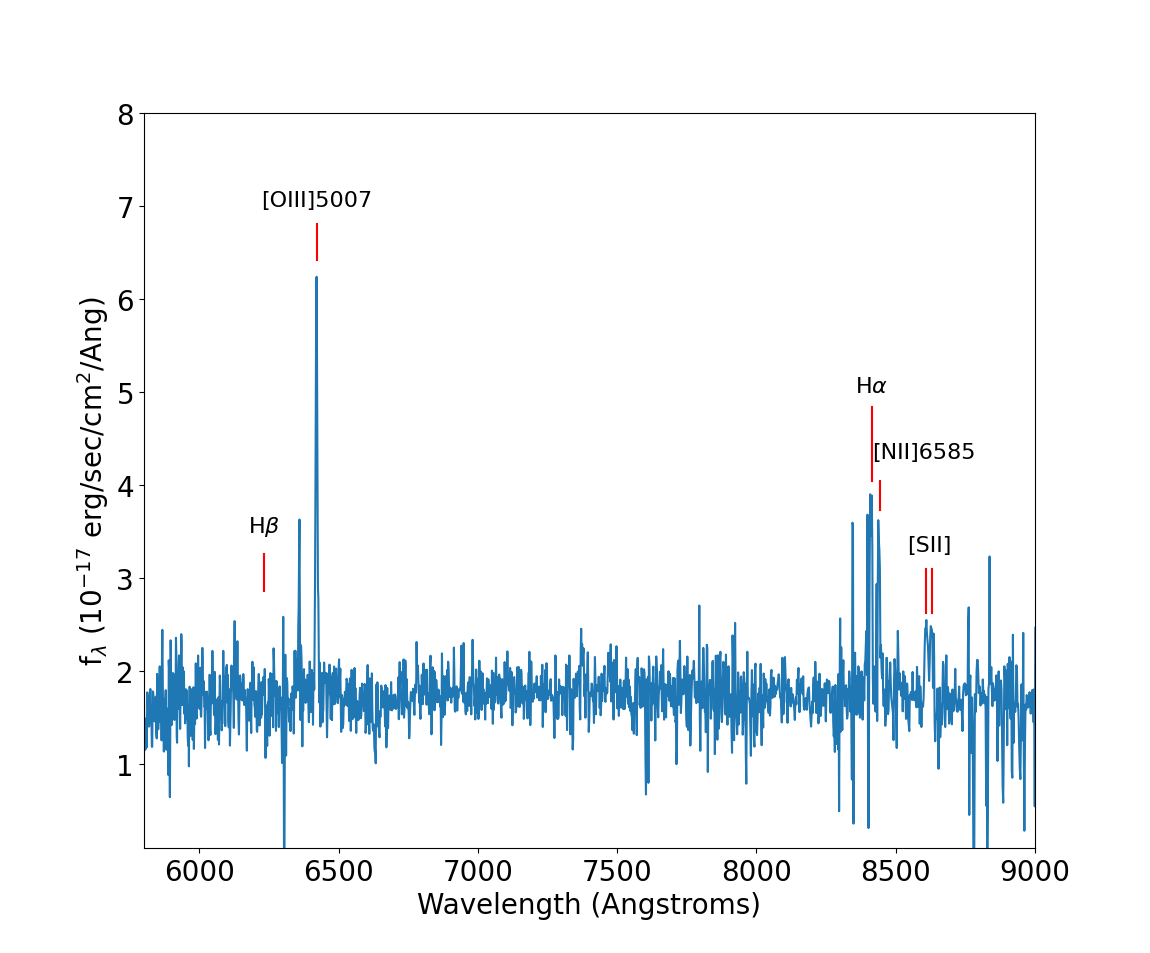}
    \caption{Top panel: Broad-line AGN spectrum (type-1 AGNs) at redshift $z=0.231$ (XLSSU J020139.1-050118). Bottom panel: Narrow-line AGN spectrum (type-2 AGNs) at redshift $z=0.282$ (XLSSU J021007.1-060459).}
    \label{fig:spectra}
\end{figure}

We classified the majority of sources with available optical spectra as either broad-line or narrow-line AGNs based on the width of Balmer emission lines, using optical spectra primarily obtained from SDSS. All spectra in the redshift range of this study include at least the H$\beta$ region, while most of them include also the H$\alpha$. In particular, we classify all sources with a FWHM of the Balmer lines less than $500$ km/s and no evidence for any broadening with respect to forbidden lines (such as [OIII], [NII]) as narrow-line AGNs. Nevertheless, a non-negligible fraction presents the typical spectrum of an absorption-line galaxy (ALG), while we were not able to reliably classify a few of them due to poor spectrum quality. In Fig.~\ref{fig:spectra} we present an example of a broad- and a narrow-line spectrum. A summary of the classification is presented in Table~\ref{table:1}, and individual classifications are given in Table~\ref{table:list}.

\subsubsection{X-ray hardness ratio}
\label{sub:HR}

The X-ray hardness ratio (HR) refers to a measure of the relative intensity of X-rays at different energies. It is typically calculated as the ratio of the counts detected in two different energy bands with the equation: 
\begin{equation}
\label{eq:2}
    HR=\frac{H-S}{H+S},
 \end{equation}
where H is the hard- and S is the soft-band count-rate. In our case, we use the [0.5-2] keV band as the soft and the [2-10] keV as the hard. The X-ray hardness ratio can serve as a proxy for absorption from the obscuring torus, assuming a spectral modelling for the continuum emission. For our purposes, we adopt a simplified model for the AGN emission, assuming a single power law fit to the data. The torus absorbs softer X-rays preferentially over harder ones due to its composition and density. Consequently, the X-ray hardness ratio, which compares the counts of X-rays at different energy bands, can indicate the degree of obscuration. High hardness ratio values indicate significant obscuration, as softer X-rays are absorbed more effectively, resulting in a relatively greater detection of harder X-rays. To compute the HR values of our sources, we used the Bayesian method described by \citet{Park2006}. 

In the current work, this method allows an estimation of obscuration for the full sample, contrary to the similar analysis using optical spectra where we only have data for approximately half of the sample. However, we note that although the X-ray and the optical obscuration are well correlated, they exhibit significant scatter \citep{Jaffarian2020} and overall, there is no a one-to-one correlation between them \citep[e.g.][also known as \citetalias{Masoura2020}]{Masoura2020}. Furthermore, there are also indications that the X-ray spectral index depends on Eddington rate \citep[e.g.][]{Bianchi2009}, and that the soft X-ray band, [0.5-2] keV, at low redshift may include contributions from the soft excess, leading to the underestimation of obscuration. Nevertheless, for the purpose of statistical comparisons in the current work, the use of HR as an indicator of obscuration is instructive. In Table~\ref{table:1} we quote the average HR values of BL and NL AGNs, and ALGs, along with their optical classification. There is a good agreement between the two obscuration proxies with the BL AGNs exhibiting significantly lower HR values than the NL population. The average HR of ALG sources does not indicate any heavy absorption in the X-ray band. In these sources, we assume that the AGN optical emission is diluted by the strong stellar continuum of their host galaxies.

\begin{table}
\caption{AGN classification}             
\label{table:1}      
\centering                         
\begin{tabular}{c c c c }          
\hline\hline                 
annulus & BL & NL  & ALG \\    
(1) & (2) & (3) & (4) \\
\hline                        
   1 & 3 (-0.50)& 6 (-0.07) & 4 (-0.55) \\      
   2 & 1 (-0.44) &  6 (0.10) & 1 (-0.76) \\
   3 - 4 & 4 (-0.66) & 12 (-0.32)  & 4 (-0.32) \\
\hline                                   
\end{tabular}
\tablefoot{(1) $r_{500}$ annulus in which the X-ray AGNs are located, (2) - (4) Number of broad-line (BL) AGNs, narrow-line (NL) AGNs, and absorption line galaxies (ALG), as classified by their optical spectra. In the parentheses we report the average HR, computed as described on Sect.~\ref{sub:HR}.}
 \end{table}

\subsubsection{Spectral Energy Distribution (SED)}
\label{sec:sed}

The AGNs (e.g. accretion power) and the host-galaxy (stellar mass and SFR) properties of the sources were derived through SED fitting techniques using the Code Investigating GALaxy Emission \textit{CIGALE} algorithm \citep{Boquien2019,Yang2020,Yang2022}. The SED fitting analysis is described in detail in \citet{Pouliasis2020,Pouliasis2022a}. In brief, we used the stellar population synthesis model defined in \citet{Bruzual2003} assuming the initial mass function by \citet{Salpeter1955} and a constant solar metallicity (Z\,=\,0.02) for the stellar emission in addition to a delayed star-formation history (with a functional form $\rm SFR\propto t\times exp(-t/\tau)$) that includes a star formation burst, no longer than $\tau = 20$ Myr \citep{Malek2018,Buat2019}. Furthermore, we used the attenuation law of the stellar emission by \citet{Charlot2000} and we modelled the dust emission of the galaxy with the templates of \citet{Dale2014} without including the AGN emission. Finally, we used the SKIRTOR model \citep{Stalevski2012,Stalevski2016} for the AGN emission at the different wavelengths without including the X-ray or radio modules The parameter space for the several modules in the SED fitting process was adopted from \citet{Mountrichas2024}. 

It is important in our analysis, to have reliable measurements of the AGN properties along with both the global M$_*$ and the SFR of the their host galaxies. For that purpose, we required our sources to have low reduced $\rm \chi^2$ ($\rm \chi^2_r$) that is indicative of the goodness of the SED fitting process. To this end, we excluded sources that have $\rm \chi^2_r > 5$ \citep[e.g.,][]{mountrichas2019,buat2021,Pouliasis2022b}.

\section{Results}
\label{sec:results}

\subsection{X-ray-detected AGNs in merging and disturbed galaxies}

\begin{figure}[]
    \centering
    \includegraphics[width=9cm]{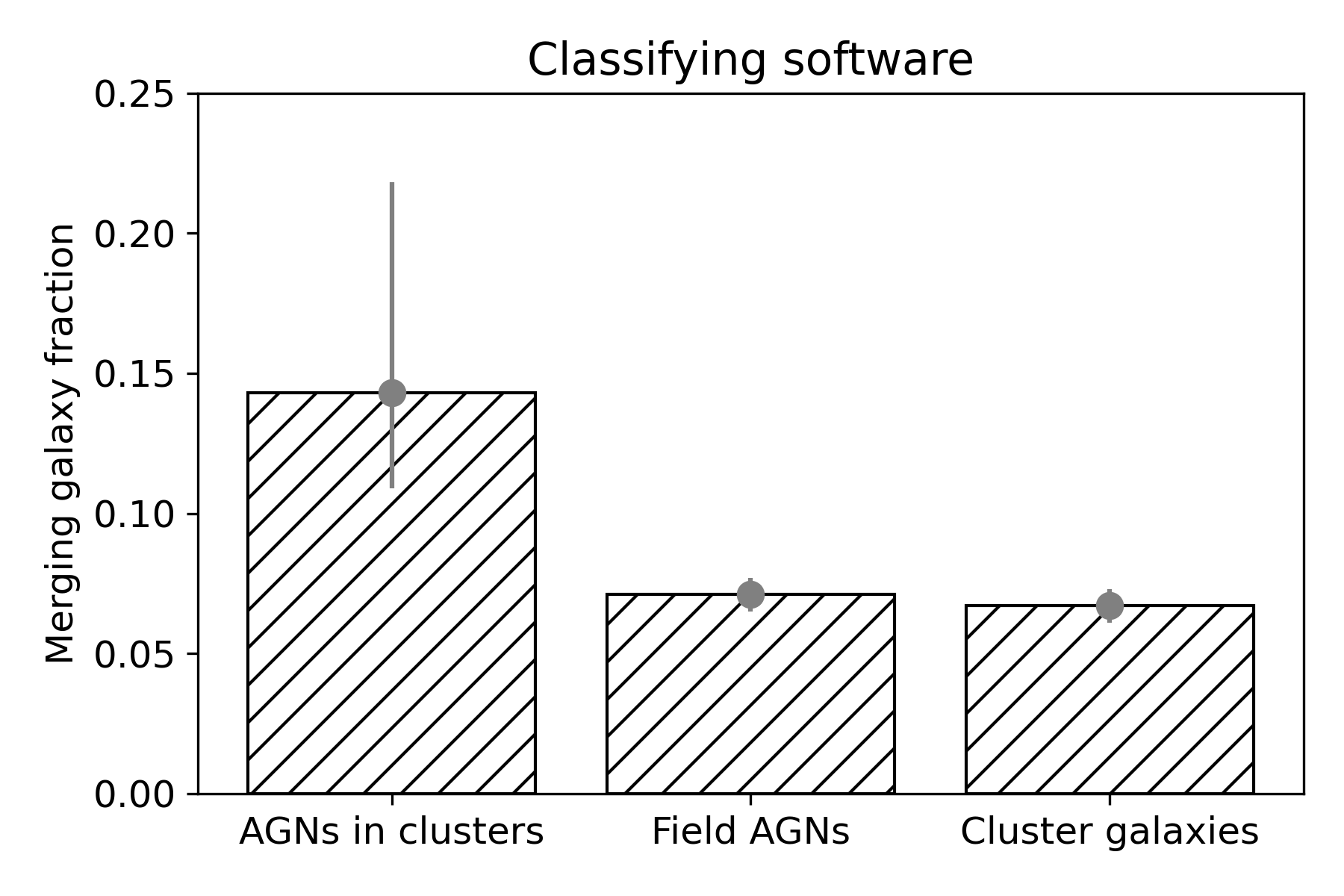}
    \includegraphics[width=9cm]{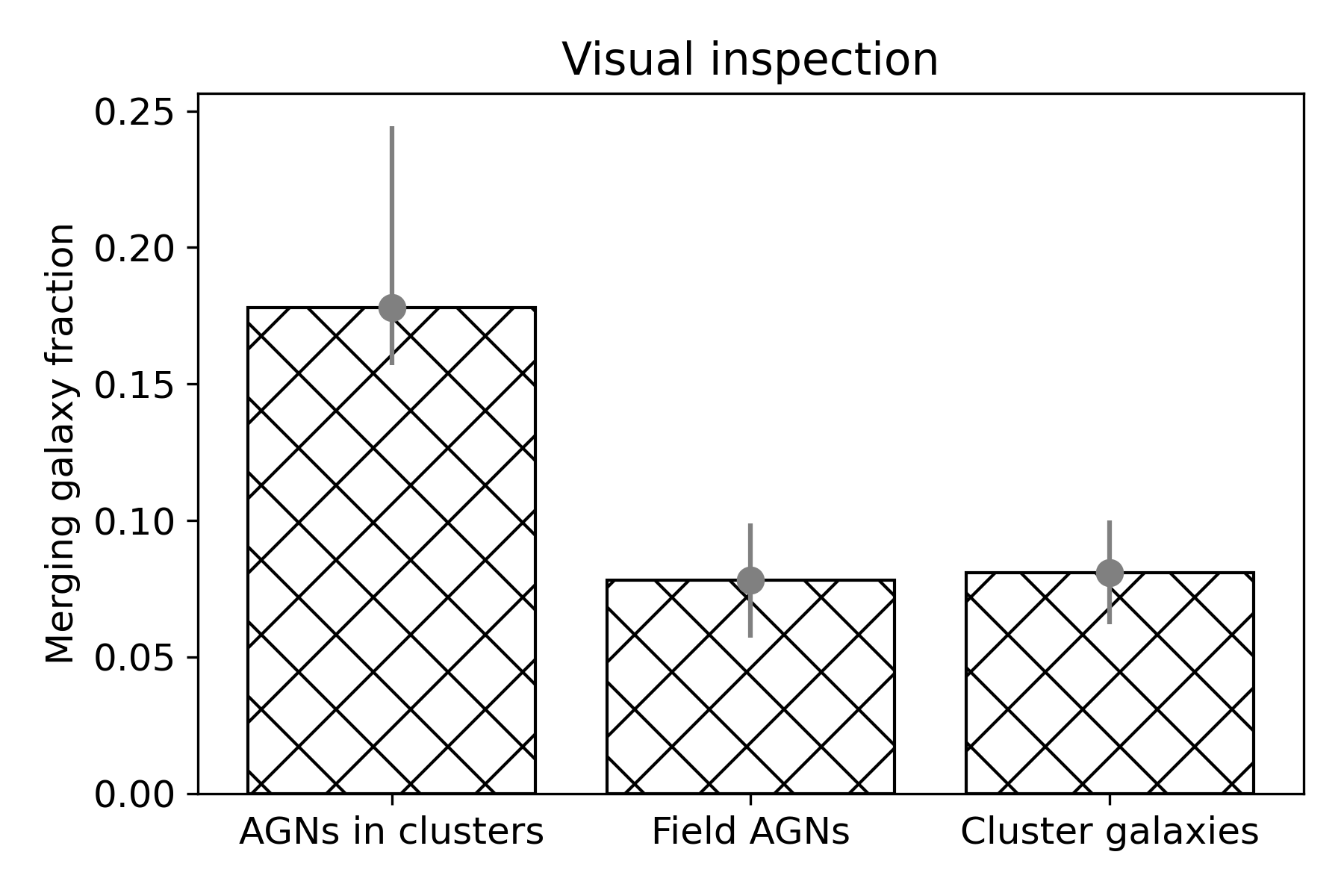}
    \caption{Fraction of merging and disturbed X-ray AGN host galaxies (main sample) in comparison with the corresponding fraction in control samples (i) and (ii) (top panel), and (iii) and (iv) (bottom panel). Error bars indicate the $ 1 \sigma $ confidence limits for small numbers of events \citep{Gehrels1986}. A significant excess of merging galaxies, at the $2\sigma$ confidence level, is found in the sample of X-ray AGNs in clusters.}
    \label{fig:smallmerg}
\end{figure}

In the current study, the merging fraction in each sample is defined as the number of merging and disturbed galaxies divided by the total number of sample galaxies. The results are presented in Fig.~\ref{fig:smallmerg}. Using either {\tt Statmorph} classifying software (see Sect.~\ref{sub:morph}) or visual classification, the merging fraction of X-ray AGN hosts within clusters is significantly higher, at the 2$\sigma$ confidence level, than in field X-ray AGNs or in non-active cluster galaxies.

\begin{figure}[htp]
    \centering
    \includegraphics[width=9cm]{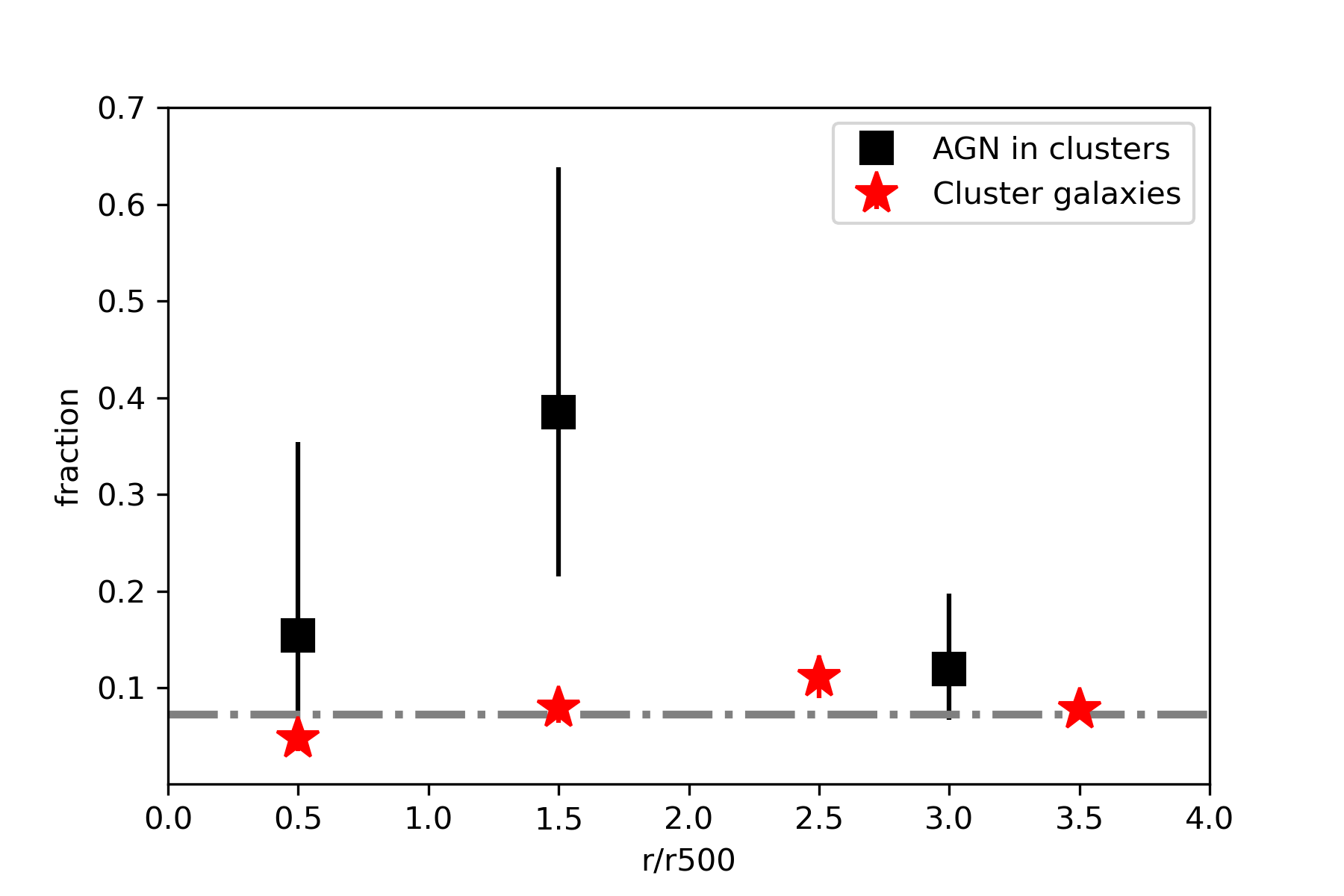}
    \caption{Fraction of merging X-ray AGNs in clusters (main sample) as a function of distance from the cluster centre. For comparison, we overlay the respective fraction for the 1914 non-AGNs cluster galaxies (control sample (ii)). The hatched line represents the average value of the merging fraction derived from the sample of 1987 field X-ray AGNs (control sample (i)). Error bars indicate the $ 1 \sigma $ confidence limits for small numbers of events \citep{Gehrels1986}. The results indicate a significant excess, at the 2$\sigma$ confidence level, of merging AGNs hosts in cluster outskirts.}
    \label{fig:clustmerg}
\end{figure}

To identify possible trends in the spatial distribution of merging hosts, we followed the methodology described in Sec.~\ref{sec:method} to segregate the area around clusters. To improve the statistics, we merge the third and the fourth annuli, which we consider to be the field. This decision was made following our initial inspection, which confirmed that the results from the two outer annuli consistently matched those of the field AGNs control samples (i and iii). Our results are plotted in Fig.~\ref{fig:clustmerg}. For comparison, we plot only the results of the large control sample of cluster galaxies (ii), as they coincide with sample (iv) but offer better statistics.

\begin{figure}[htp]
    \centering
    \includegraphics[width=9cm]{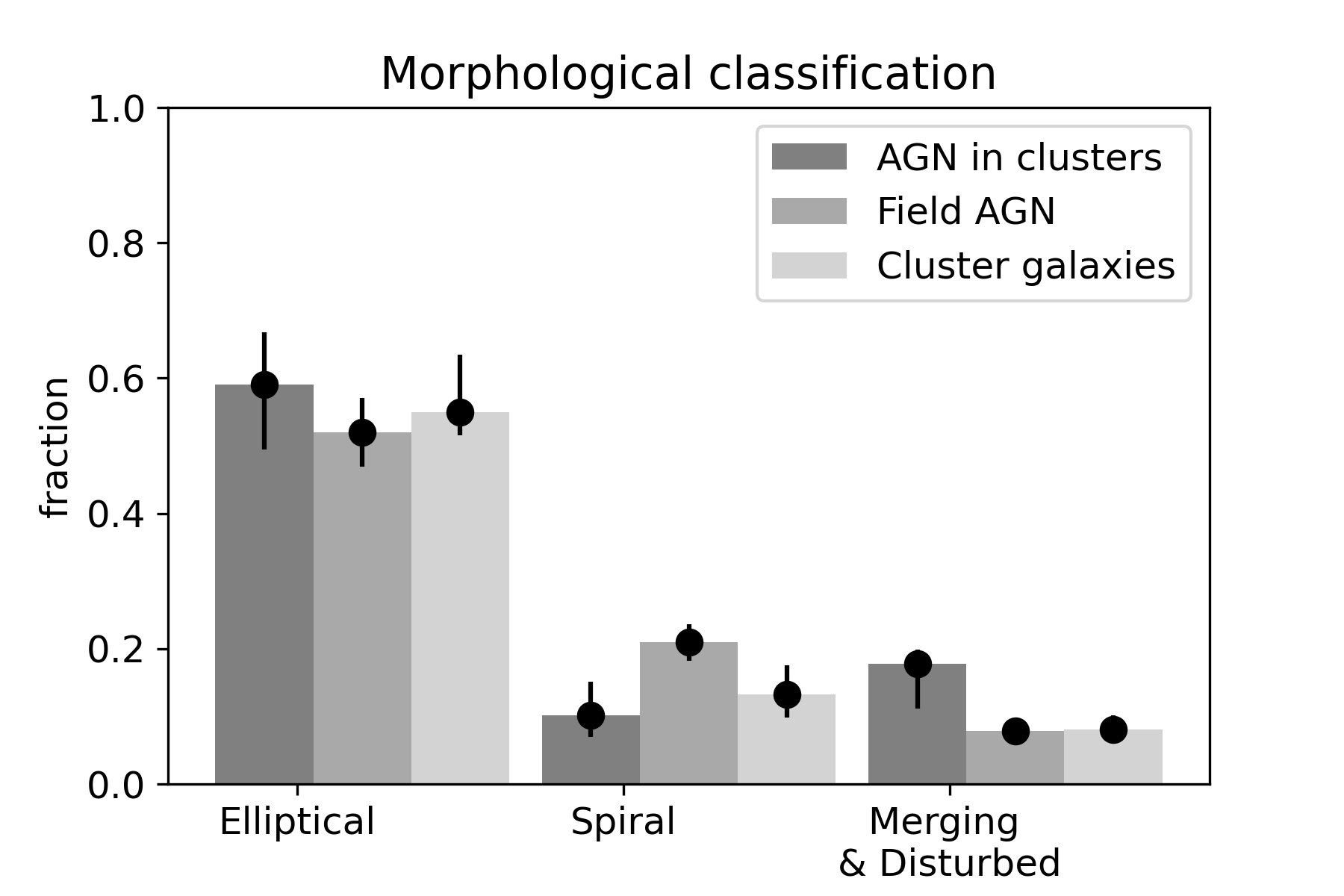}
    \caption{Visual morphological classification of X-ray AGN hosts in clusters in comparison with field X-ray AGNs (iii) and non-AGN cluster galaxies (iv). The classification was based on visual inspection of HSC images. Dominance of elliptical shapes was expected, particularly among cluster galaxies. The careful selection of control-sample galaxies based on redshift and stellar mass ensured the inclusion of similar galaxy types across all samples. Error bars indicate the $ 1 \sigma $ confidence limits for small numbers of events
    \citep{Gehrels1986}.}
    \label{fig:morph}
\end{figure}

Our analysis revealed that the majority of merging and disturbed X-ray AGN hosts reside in cluster outskirts. In particular, their fraction in the second annulus is  significantly higher, at the 2$\sigma$ confidence level, than the respective fraction in non-active cluster galaxies and field X-ray AGNs. The merging fraction in the two outer annuli ($0.12^{+0.08}_{-0.05}$) is consistent with the corresponding fraction ($0.071\pm 0.006$) in the control samples. This confirms our assumption that beyond the 2$r_{500}$ radius, AGNs behave similarly to the field population, as previously reported in \citetalias{Koulouridis2018b} for the XXL sample below $z<0.5$. However, the difference between the outskirts and the cluster centre is not significant at any confidence level, due to the small number of sources. The statistical results of the visual classification, including spiral and elliptical hosts, are presented in Fig.~\ref{fig:morph}, while individual classification and images in Table~\ref{table:list} and Table~\ref{table:images}, respectively.

\begin{figure}[h]
\centering
\includegraphics[width=9cm]{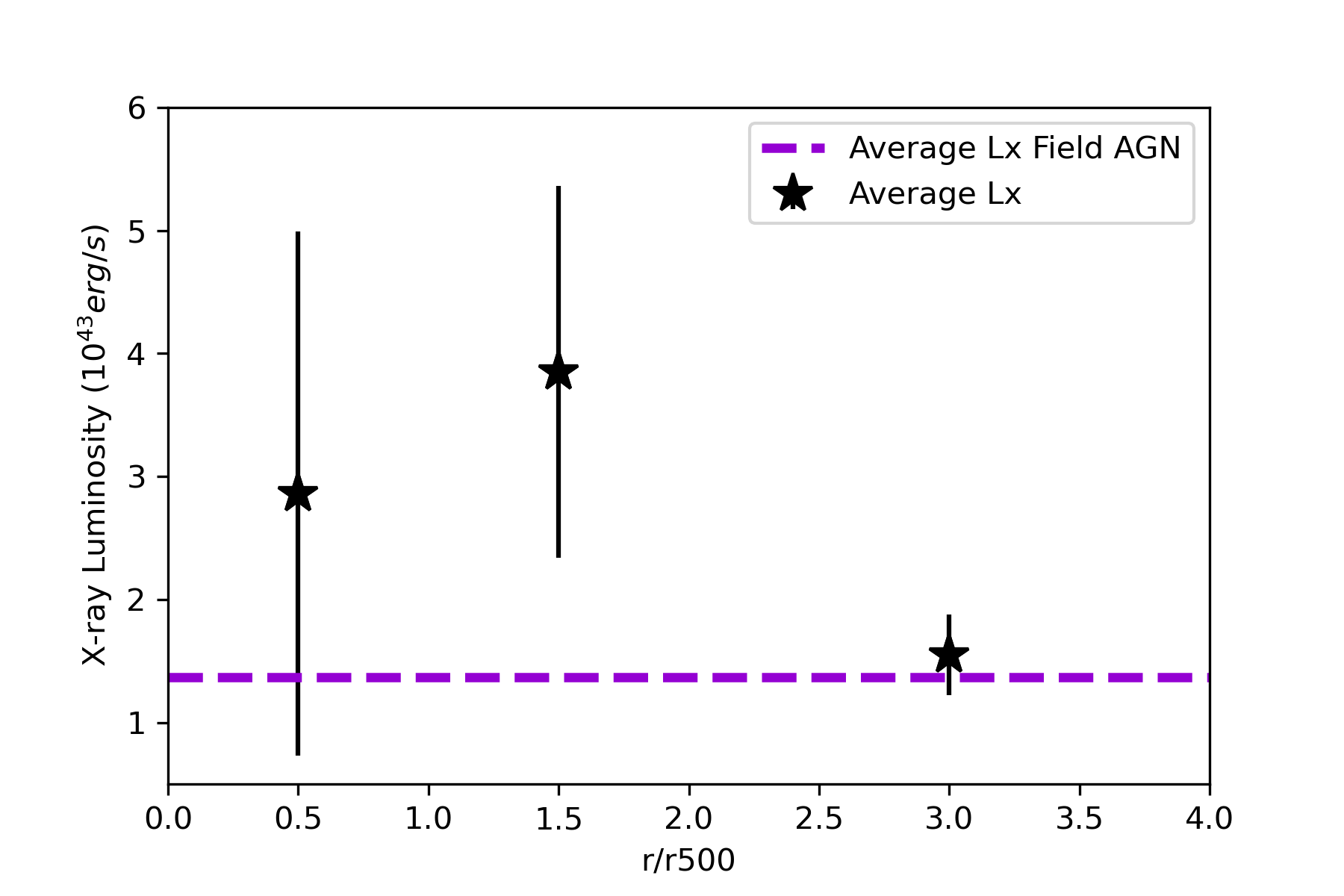}
    \caption{Average X-ray luminosity (Lx$_{[0.5-10]\;\rm keV}$) of X-ray AGNs in clusters. The hatched line represents the average value of Lx derived from the sample of 1987 field X-ray AGNs (control sample (i)). Error bars denote the $1\sigma$ confidence limits.
   }
    \label{fig:Lx}
\end{figure}

 Finally, in Fig.~\ref{fig:Lx} we plot the average X-ray luminosity (Lx$_{[0.5-10]\;\rm keV}$) of AGNs within clusters, which is a direct output of the XXL pipeline as described in \citet{Faccioli2018}, also known as \citetalias{Faccioli2018}. It highlights how their radiative output varies with distance from the cluster centre. Similarly to the merging fraction, the average X-ray luminosity is three times higher than in the field, at the $\sim1.5\sigma$ confidence level, while the outer annuli are in agreement with the field value. This may further support the idea that the relatively high number of galaxy mergers in the cluster outskirts contribute to the triggering of AGNs.

\subsection{AGN Obscuration}

In this section, we examine the obscuration of X-ray AGNs in our samples. Our aim was to investigate potential trends that may reveal the physical conditions responsible for triggering AGNs in the dense environments of galaxy clusters. We use two indicators of obscuration, as described in Sect.~\ref{sec:method}; the AGN optical type (type-1/type-2) and the X-ray hardness ratio.

\subsubsection{Optical AGN classification}
\label{sub:opt_res}
 
\begin{figure}[htp]
    \centering
    \includegraphics[width=9cm]{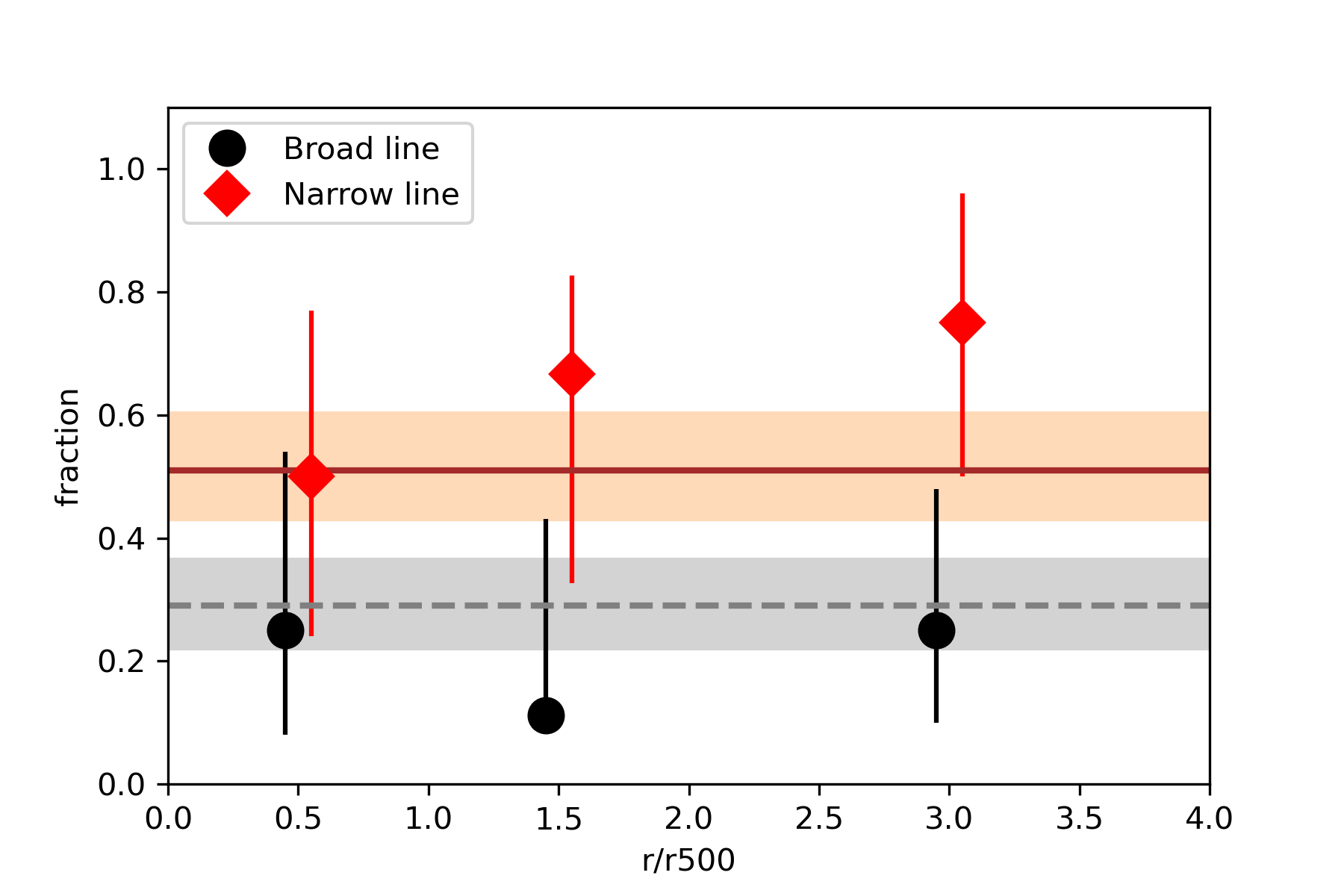}
    \caption{AGN optical classification of X-ray AGNs in clusters (black circles for broad-line and red squares for narrow-line AGNs) compared to the control sample of field X-ray AGNs (dashed black line for broad-line and continuous red line for narrow-line AGNs). The classification and detailed results are described in Sect.~\ref{sub:optical} and Sect.~\ref{sub:opt_res}. 
    Error bars and shaded areas indicate the $ 1 \sigma $ confidence limits for small numbers of events \citep{Gehrels1986}.}
    \label{fig:spectype}
\end{figure}

First, we investigate potential trends related to AGN classification based on the optical spectra of our X-ray-detected AGNs. In total, we find eight broad-line and 22 narrow-line AGNs. Another six sources have an absorption-line galaxy (ALG) spectrum, typical of elliptical galaxies. However, most of the ALG present some weak emission lines, especially in the H$\alpha$ region. Optical emission from these AGNs, which are hosted mostly by massive elliptical galaxies, could be diluted by the stellar continuum. In addition, these AGNs may be intrinsically weak, since the X-ray luminosity for four out of six is below $3\times10^{42}$ erg/sec, placing them in the first quartile. 
The results are presented in Fig.~\ref{fig:spectype}. We were unable to identify any significant trends in the cluster-centric radial distribution of optical AGN types, likely due to the small sample size and resulting large uncertainties. 

The ratio of type-2 to type-1 AGNs identified in the current study is consistent with the field ratio observed in the Local Universe \citep[e.g.][]{Maia2003} and in high-redshift clusters ($z\sim1$) \citep{Mo18}. This is in good agreement with a similar study of spectroscopically confirmed X-ray AGNs in 19 galaxy clusters \citep{Koulouridis2024}, within a narrow redshift range $0.16 \leq z \leq 0.28$. However, our findings diverge from recent results obtained from local clusters in the WINGS and Omega-WINGS surveys. In those studies, a notably higher optical type-2 to type-1 fraction was reported (approximately 10 to 1) compared to the field \citep{Marziani2023}. However, the sample selections in these studies differ significantly, so any comparisons should be approached with caution.

Interestingly, \citet{Koulouridis2024} reported that three out of the four broad-line AGNs within $2r_{500}$ were found in the central $r_{500}$ annulus. They argued that close to the cluster core, AGN activity may be triggered either by the influence of strong RPS, as suggested by studies of "jellyfish" galaxies \citep{Poggianti2017, Peluso2022}, or by tidal shocks, as galaxies pass through cluster pericentre. Our results support these findings, since three out of the four broad-line sources up to $2r_{500}$ were found again in the same region. We note, that the cluster mass distribution of the current sample, mostly comprising groups and poor clusters, is markedly different from the massive clusters used in \citet{Koulouridis2024}.  
In addition, the SED analysis showed that the accretion power of these three broad-line sources is higher than that of the rest of the X-ray AGNs sample, and also of the other five broad-line AGNs found farther from the cluster centre. We note that the fraction of broad-line AGNs in the first two annuli is comparable to that in the last two annuli, which represent the field. However, the environment is drastically different.

We note that only half of our X-ray AGNs have available spectroscopic data. Consequently, in the next section, we present the results of a statistical analysis based on the X-ray hardness ratio of our entire source sample.

\subsubsection{X-ray hardness ratio}
\label{sub:HR_res}

\begin{figure}[htp]
    \centering
    \includegraphics[width=9cm]{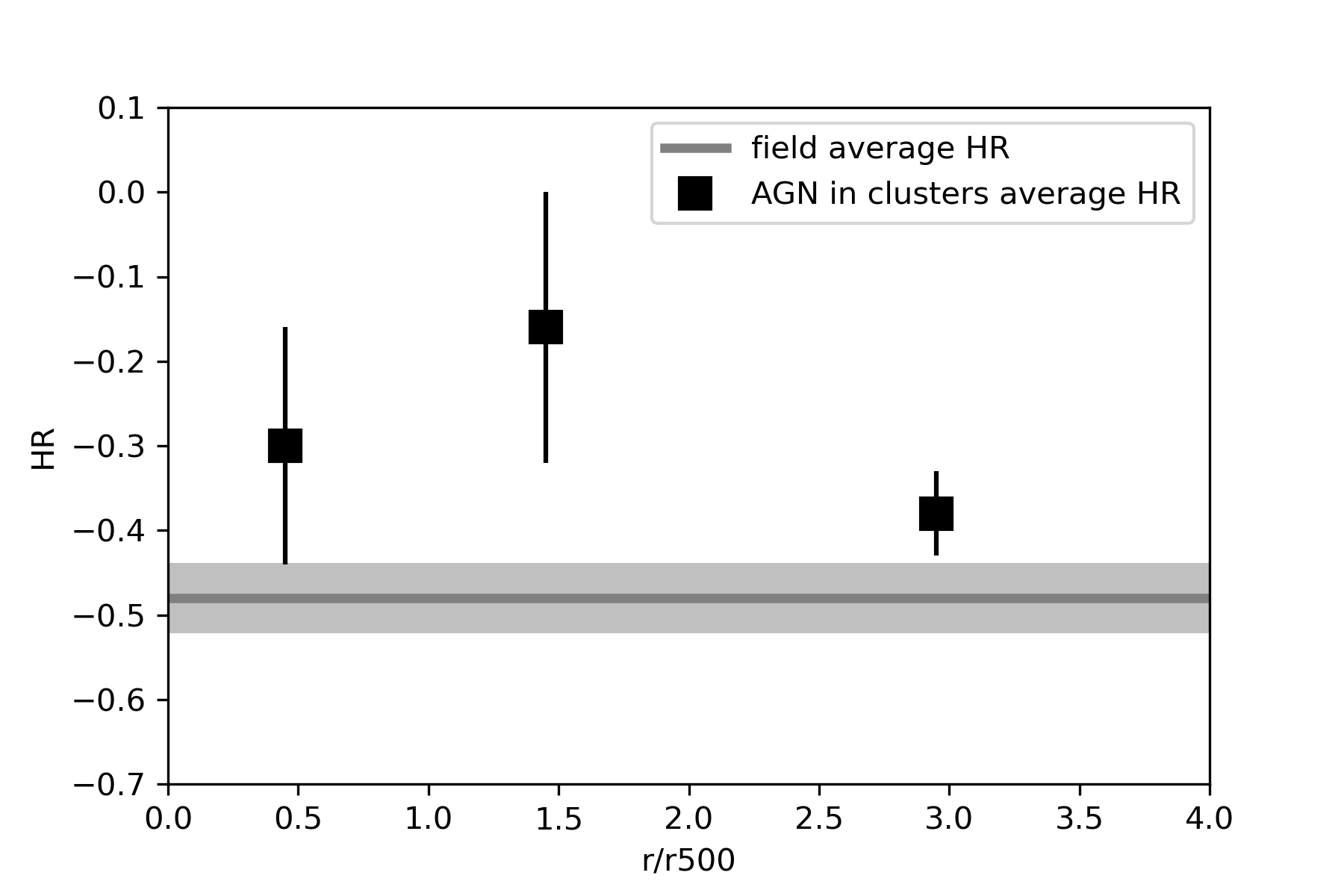}
    \includegraphics[width=9cm]{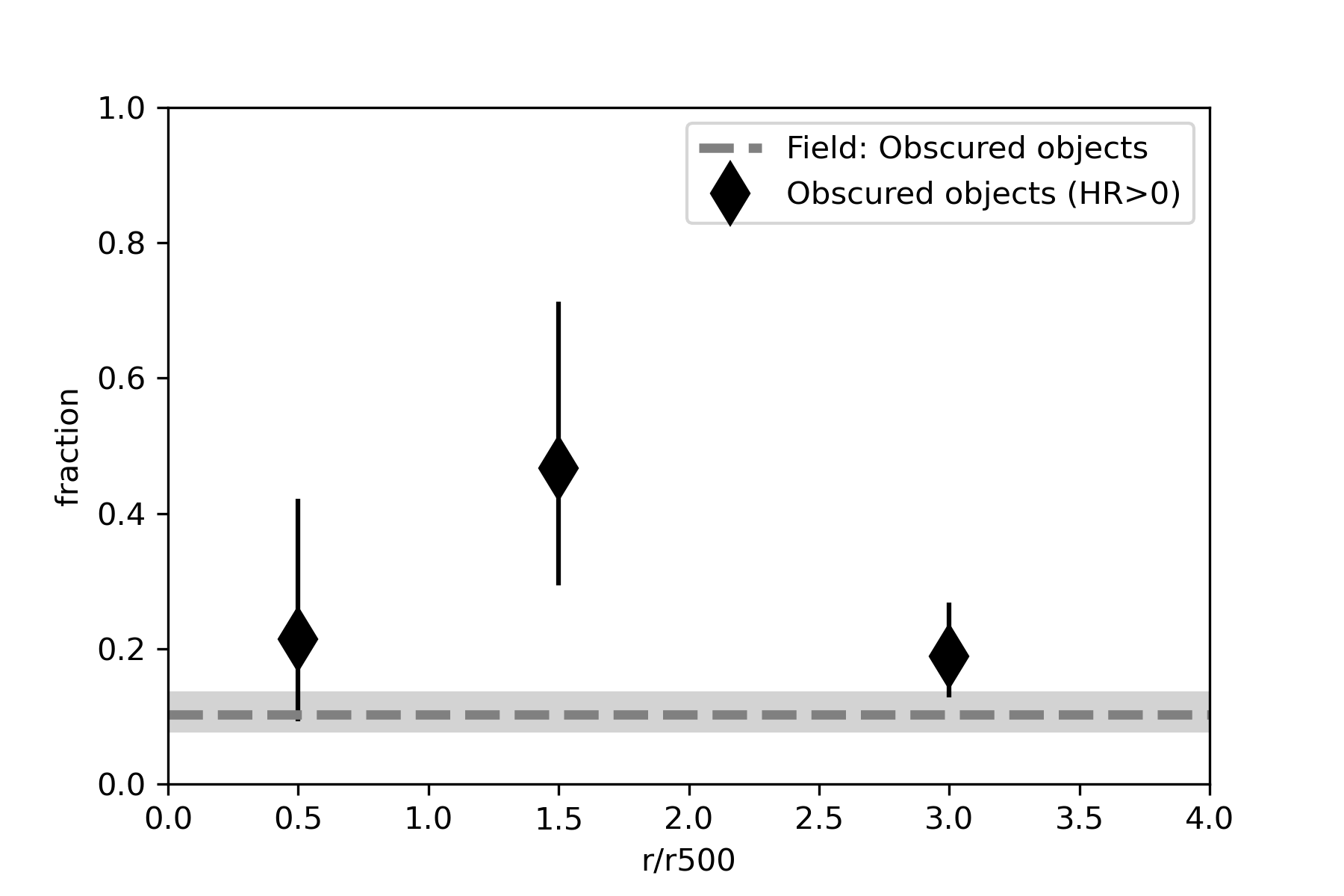}
    \caption{Top panel: Average hardness ratio (black squares) of X-ray detected AGNs in clusters divided in three cluster-centric annuli, 0-1R$_{500}$, 1-2R$_{500}$ and 2-4R$_{500}$ with their standard $1 \sigma $ errors. The grey line corresponds to the average hardness ratio value of field AGNs (the shaded area denotes the $1\sigma$ error range). Evidently, the average HR in the 1-2R$_{500}$ area is significantly higher than in the field, at the 90\% confidence level \citep{Gehrels1986}. Bottom panel: Obscured X-ray AGN fraction in clusters. We consider as obscured AGNs the objects with HR>0 (see Sect.~\ref{sub:HR_res}). The dashed line represents the fraction of the obscured X-ray AGNs in the field (the shaded area denotes the $1\sigma$ error range). There is a significant excess of X-ray AGNs in cluster  outskirts, at the $2 \sigma$ confidence level.}
    \label{fig:HR}
\end{figure}
The X-ray hardness ratios of cluster and field X-ray AGNs were computed following the methodology described in Sect.~\ref{sub:HR}. For our purposes, we then derived the mean HR in each cluster annulus. The results are presented in Fig.~\ref{fig:HR}. We cannot claim any difference between the outskirts and the central annulus because of the large uncertainties. Nevertheless, The mean HR values indicate that X-ray AGNs in cluster outskirts are more obscured compared to those in the outer annuli (2-4$r_{500}$), at the 1$\sigma$ confidence level, or in the field, at the 90\% confidence level \citep{Gehrels1986}.

The higher average obscuration observed in the outskirts compared to the field may result from a higher number of obscured sources or, alternatively, from the presence of a few heavily obscured sources. To determine the source of obscuration, we needed to estimate the fraction of sources that are obscured in each annulus. To this end, we used PIMMS online tool\footnote{https://heasarc.gsfc.nasa.gov/cgi-bin/Tools/w3pimms/w3pimms.pl} to compute the HR value that would best define a threshold between obscured and unobscured sources based on the column density of the obscuring material. A reasonable value of the column density above which we can assume that a source is obscured is $10^{22}$ cm$^{-2}$. The HR that corresponds to this value depends on redshift and the slope ($\Gamma$) of the X-ray spectrum. Our computations within the redshift range of our sources and for various values of $\Gamma$ result in an average HR=0 as a reasonable threshold. The results are plotted in the bottom panel of Fig.~\ref{fig:HR}. The fraction of obscured sources in the outskirts of clusters are significantly higher than the corresponding fraction in the field at the 2$\sigma$ confidence level. However, we cannot confirm any statistically significant difference between the outskirts and the centre of clusters due to small number statistics.

The results from the HR analysis are seemingly not in agreement with the results from the optical spectra analysis presented in the previous section. However, the two analyses are not directly comparable, since the samples are different. The current sample includes many faint sources (m$_i$>21) without spectroscopic data, possibly more obscured than the spectroscopic part of the sample. More importantly, as described in Sect.~\ref{sub:HR}, the correlation between these two obscuration proxies presents large scatter. 

In summary, the average obscuration is higher in cluster outskirts than in the field, at the $2 \sigma$ confidence level, due to a larger number of obscured AGNs. The obscuration excess coincides with the merging excess shown in Fig.~\ref{fig:clustmerg}, further supporting galaxy merging as the main AGN triggering mechanism in the outskirts.

\subsection{AGNs and host galaxy properties derived from SED fitting}

The SED analysis of the X-ray AGNs in our cluster sample may reveal some interesting environmental trends. Specifically, we statistically examine the black hole mass ($M_{BH}$), the black hole accretion rate (BHAR), and the Eddington rate ($\lambda_{edd}$) of the AGNs, as well as the SFR of their host galaxies. 

In order to estimate the $M_{BH}$, we used the following equation
from \citet[][]{Kormendy2013} that takes into account the stellar-mass $(M_{*})$ dependence:
\begin{equation}
M_{BH} = 4.9 \times 10^{-3} \left( \frac{M_{*}}{10^{11} M_\odot} \right)^{0.14}M_{*},
\end{equation}
where $M_{*}$ is the quantity "bayes.stellar.mstar" from CIGALE. The average BHAR and $\lambda_{edd}$ may provide insight into the mass growth rates of black holes across different cluster regions. For the calculation, we used the following equation from \citet[][]{yang2023,pouliasis24}:
\begin{equation}
    \text{BHAR} = \frac{L_{\text{disk}} (1 - \varepsilon)}{\varepsilon c^2} = 1.59 \times \frac{L_{\text{disk}}}{10^{46} \, \text{erg s}^{-1}} \, [M_{\odot} \, \text{yr}^{-1}],
\end{equation}
where $L_{\rm disk}$ is the quantity "AGN.accretion.power" from our SED fits, which is the viewing-angle-averaged intrinsic accretion-disk luminosity. Also, $\varepsilon$ is the radiative efficiency, the fraction of the accreted mass converted into radiation. For comparison, we derive the $\lambda_{edd}$ of sources both from the SEDs and from the X-rays, assuming $L_{\text{disk}} \sim L_{\text{bol}}$. In particular, the $\lambda_{edd}$ calculated from the SEDs is: 
\begin{equation}
 \lambda_{edd} = \frac{L_{\text{disk}}}{L_{edd}},
\end{equation}
while from the X-rays it is
\begin{equation}
 \lambda_{edd} = \frac{{Kx(Lx)\times Lx}}{L_{edd}},
\end{equation}
where $L_{edd}=1.3\times10^{38}(M_{BH}/M_\odot)$ and $Kx(Lx)$ is the bolometric correction as described in \citet{Duras2020}.

\begin{figure*}[h]
\centering
\includegraphics[width=0.42\textwidth]{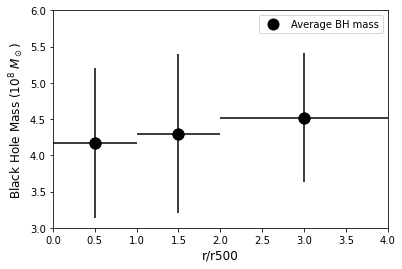}\includegraphics[width=0.42\textwidth]{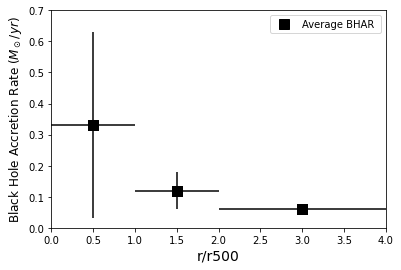}
\includegraphics[width=0.42\textwidth]{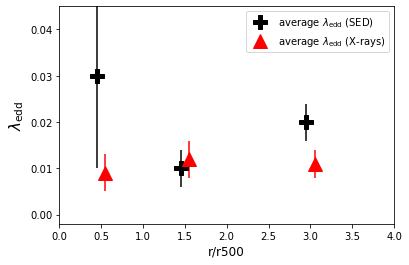}\includegraphics[width=0.42\textwidth]{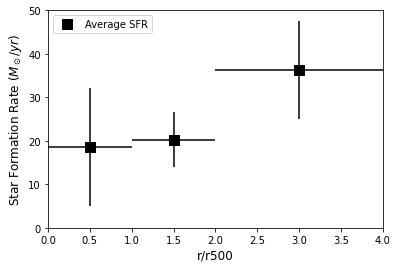}
    \caption{Summary plot of X-ray AGNs and host galaxy properties as a function of cluster-centric distance. From top left to bottom right panel: Average $M_{BH}$, BHAR, $\lambda_{\text{Edd}}$ of AGNs, and average SFR of their host galaxies. Error bars denote the $1\sigma$ confidence limits.
   }
    \label{fig:BH}
\end{figure*}

The top panels of Fig.~\ref{fig:BH} illustrate the average values of the $M_{BH}$ and the BHAR as a function of cluster-centric distance. We found an average $M_{BH}\sim4.5\times 10^8 M_\odot$ and BHAR $\sim10^{-1} M_\odot/yr$ for all X-ray AGNs in all annuli. Similarly, we find no statistically significant differences in the Eddington rate of AGNs in clusters (bottom left panel of Fig.~\ref{fig:BH}), either calculated from the SEDs or directly from the X-rays. However, we found evidence that the average SFR (bottom right panel of Fig.~\ref{fig:BH}) is decreasing toward the cluster centres, although the uncertainties are large. This is in agreement with results from both the local and high-redshift Universe, where the mean SFR was always found to be higher in field galaxies than in the cluster centres \citep{treu3, poggianti2006, raichoor, Haines2015}. The decreasing SFR is probably an indication of the ram pressure induced by the ICM and frequent galaxy-galaxy interactions, which are very effective in stripping the gas of infalling galaxies \citep[e.g.][]{lars1980, cole2000, balogh2000, kawata2, Boselli2022}, transforming them into inactive ellipticals. 

In addition, sources closer to cluster centres have slightly larger stellar masses ($\rm med(M_{*}) \approx 6.5 \times10^{10}~ M_\odot$) compared to AGNs in the 2-4 annuli ($\rm med(M_{*}) \approx 5.1 \times10^{10}~ M_\odot$). Further analysis of the SFR-stellar mass relation of our AGNs relative to the main sequence (MS) of the star-forming galaxies is presented in Appendix~\ref{app:SFR}.
The SEDs plotted in Fig.~\ref{fig:SEDs} are examples (with a good fit of $x^2 < 5$) of different categories of host galaxies of our sample, namely a spiral, an elliptical, and a merging host galaxy, respectively.

\begin{figure}[htp]
    \centering
    \includegraphics[width=0.47\textwidth]{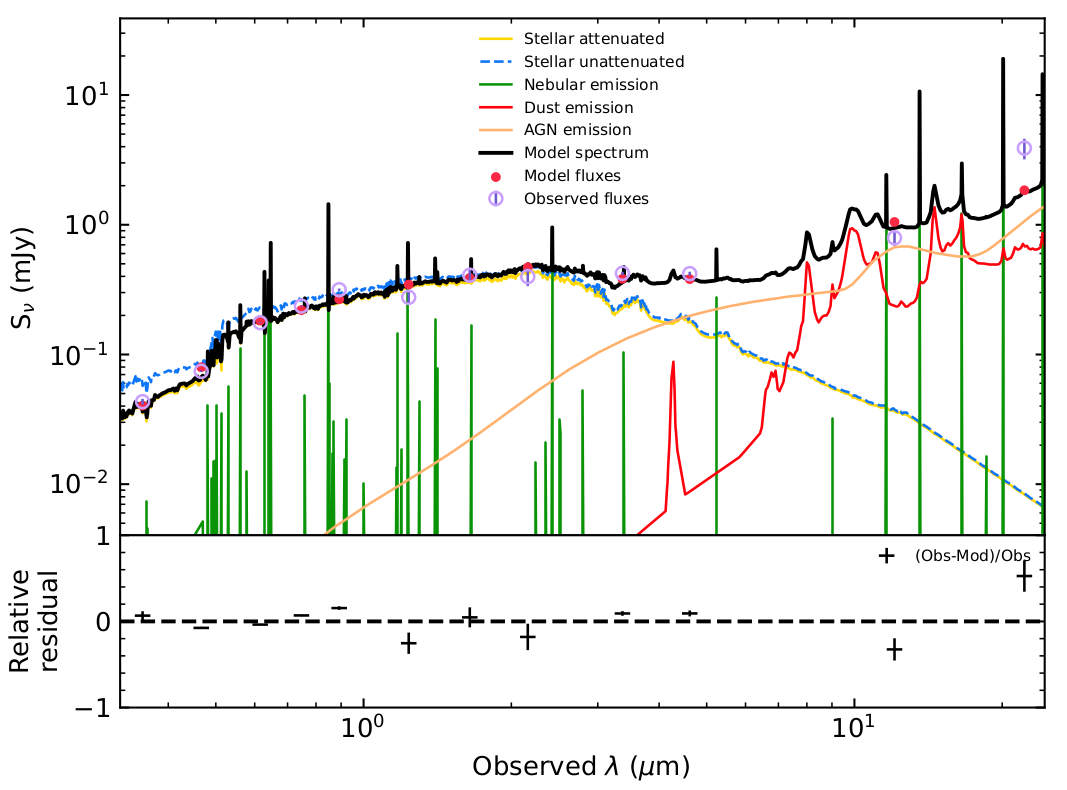}
    \includegraphics[width=0.47\textwidth]{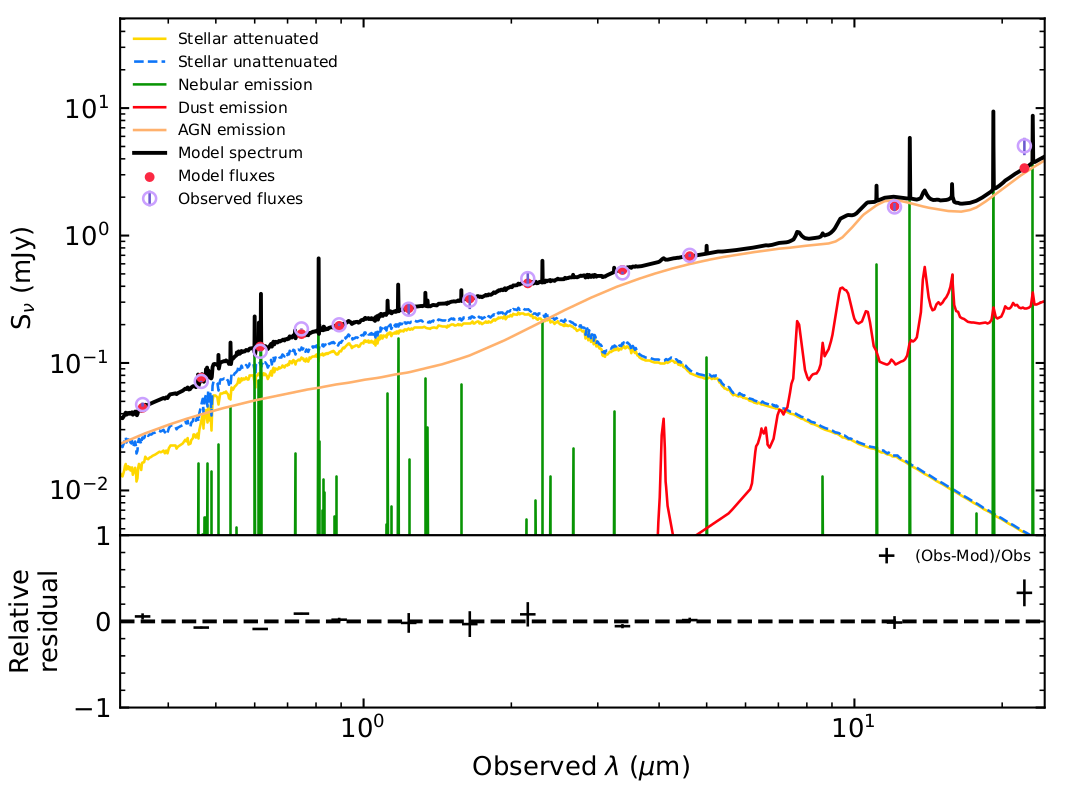}
    \includegraphics[width=0.47\textwidth]{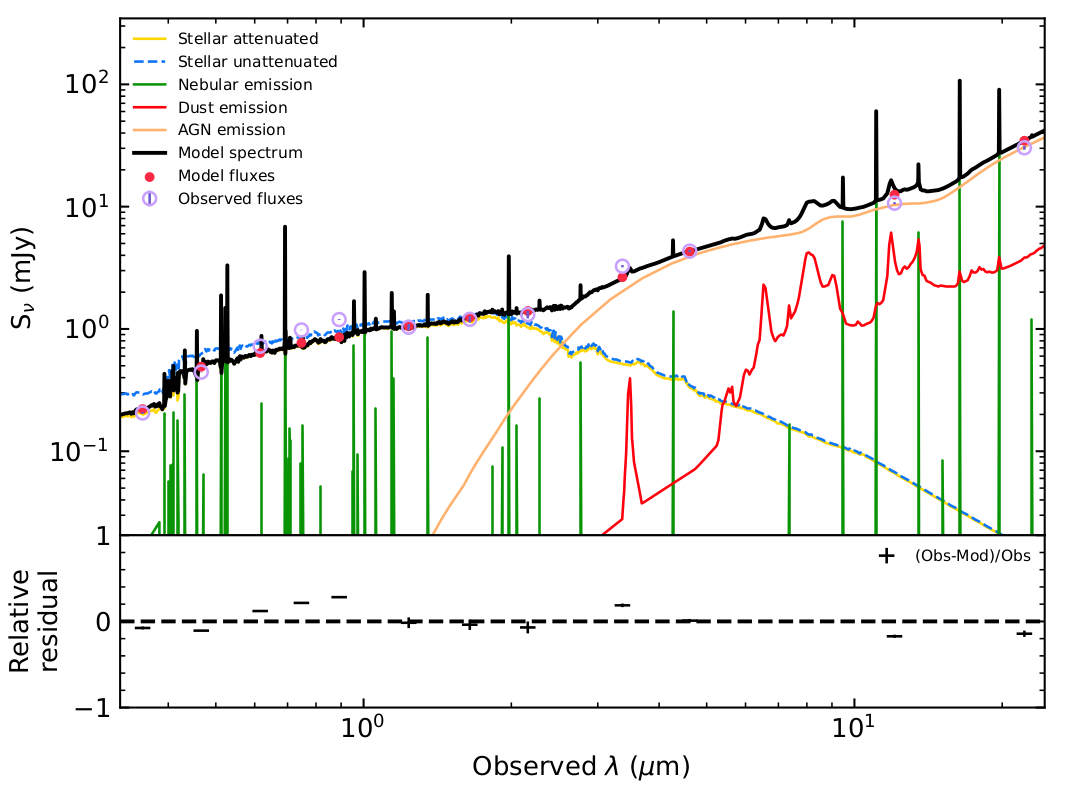}
    \caption{SED examples of AGN host galaxies. Top panel: Broad-line passive/disturbed spiral galaxy at $z$=0.445 (4XLSSU J021118.8-042516). Middle panel: Narrow-line elliptical galaxy at $z$=0.054 (4XLSSU J022700.7-042022). Bottom panel: Narrow-line merging galaxy at $z$=0.959 (4XLSSU J021616.8-045033)}
    \label{fig:SEDs}
\end{figure}

\section{Discussion and conclusions}
\label{sec:disc}

In the current study, our aim was at investigating the role of specific properties of X-ray-detected AGNs in galaxy clusters, such as the merging fraction and the morphology of their host galaxies, obscuration, star formation and accretion rate. To this end, we have used a main sample of 82 X-ray AGNs detected within 4$r_{500}$ radii of XXL clusters. Moreover, we assembled four control samples (consisting of field X-ray AGNs and non-AGN cluster galaxies) to enable a thorough comparative analysis.

We discovered that cluster galaxies that host X-ray AGNs are significantly more often merging or disturbed, at the $2\sigma$ confidence level, than similar X-ray AGN hosts in the field and cluster galaxies without X-ray detected AGNs. These findings align with earlier research, which indicates that the interactions occurring during mergers may play a crucial role in triggering the formation of AGNs in dense environments \citep[e.g.][]{Ehlert15,Koulouridis2024}. Mergers can trigger the formation of new stars, as well as the recirculation of gas and dust. Furthermore, they are thought to be a very efficient way to trigger AGN activity, as they can provide a large amount of gas and dust to fuel the SMBH \citep[e.g.][]{Silk1998,Granato2004,Haring2004,Springel2005,Koulouridis2006a,Hopkins2008a,Zubovas2012,Steffen2023,Li2023,Comerford2024,Duan2024,LaMarca2024,Bonaventura2025}. 

Likewise, we found that the excess of merging and disturbed AGN hosts can be largely attributed to galaxies in cluster outskirts (between $r_{500}$ and 2$r_{500}$). This result supports the scenario in which the excess of AGN activity in the outskirts of clusters \citep{Johnson2003,Ruderman2005,branchesi2007,Fassbender2012,Koulouridis14,Koulouridis2019,Koulouridis2024,Hashiguchi} can be attributed to a high frequency of galaxy interactions and merging. While high-velocity dispersions in massive clusters may diminish the probability of galaxy interactions \citep[e.g.][]{Arnold09,Manzer2014}, \citet{Haines12} proposed the existence of two separate groups of X-ray AGNs in the outskirts: newly infalling galaxies and those that have already passed the pericentre and are nearing the apocentre of their orbit, termed the back-splash population. These galaxies likely have low velocities, potentially facilitating interactions and mergers with other galaxies in the outskirts. Supporting our findings, a recent study of mergers in six low-redshift clusters suggested that post-merging galaxies likely merged in low-velocity environments, such as clusters outskirts and dynamically relaxed clusters \citet{Kim2024}. 

A high merging frequency may result in increased absorption of the AGN emission, as obscuring material is driven towards the galaxy nucleus \citep[e.g.][]{Hopkins2008a,Koulouridis14,Koulouridis16a,LaMarca2024}. Therefore, investigating the obscuration of the sources may provide some further evidence of the mechanisms that trigger AGNs in clusters. To this end, we have followed two different approaches to estimate the intrinsic absorption in the samples under study: investigating (a) the optical spectra of the sources that have spectroscopic data ($\sim$50\% of the sample), and (ii) the X-ray hardness ratio (full sample). The classification of the optical spectra showed a predominance of narrow emission-line spectra among AGNs situated in galaxy clusters. However, the broad to narrow-line ratio in clusters (up to 2R$_{500}$) is not dissimilar to the respective ratio in the field. Nevertheless, three out of four broad-line sources are located within the central R$_{500}$ annulus, including the only powerful QSO (with AP $=5\times10^{44}$ erg/sec, $L_X= 3\times10^{44}$ erg/sec and no visible host galaxy) found in this sample. Despite the small number statistics, this is an interesting trend corroborated by recent similar results from a different sample of 19 clusters below $z=0.5$ \citep{Koulouridis2024}. As suggested in that publication, this trend may indicate a different triggering mechanism or accretion rate, between the cluster's core and its periphery. In more detail, AGN activity near the cluster centre might arise from two primary triggers: the influence of strong RPS, as indicated by research on "jellyfish" galaxies \citep{Poggianti2017, Peluso2022}, and tidal shocks occurring as galaxies traverse the pericentre of a cluster. It is important to note that the SED analysis revealed a significantly higher value of accretion power in these broad-line AGNs. This indicates that the processes driving the growth of these SMBH are more intense, providing further support for the AGN triggering by RPS scenario. The accretion power rate is a crucial factor in understanding the dynamics and evolution of these AGNs and influences the emission and the overall energetic output in cluster cores. This heightened activity within the core region suggests that a number of centrally located AGNs may have more efficient energy feedback, which could have significant implications for the surrounding cluster environment.

Regarding the X-ray hardness ratio distribution of the full sample, our analysis indicates that the emission coming from AGNs located in cluster outskirts is significantly more absorbed when compared either to the control sample of field X-ray AGNs (at the 90\% confidence level) or to the 2-4R$_{500}$ region (at the $2\sigma$ level). Furthermore, we confirmed that this higher obscuration is due to a high number of obscured sources in the outskirts and not just a few highly obscured ones. These findings align with the noted increase in galaxy merging and interactions within the outskirts. The excess of obscured sources in this region suggests that the dense, dynamic environment, characterised by frequent merging events, plays a crucial role in the process of AGN triggering.

\section{Summary}
\label{sec:conc}

The conclusions that can be drawn from the above analysis are the following:

   \begin{itemize}
     \item We identified a significant excess, at the 2$\sigma$ confidence level, of X-ray AGNs in galaxy clusters hosted by merging or disturbed galaxies, compared to non-active cluster galaxies or X-ray AGNs in the field. This excess can be localised in the cluster outskirts (between 1 and $2r_{500}$ radius). Previous studies suggest a connection between galaxy interactions and the triggering of AGNs in cluster environments. We argue that galaxy merging and interactions are more likely to happen in the outskirts of clusters, causing a higher frequency of AGN triggering than in field galaxies and thus, leading to the observed excess.  

    \item The hardness ratio distribution indicates higher level of obscuration in clusters than in the field, specifically in the outskirts, further supporting that galaxy merging and interactions may act as an AGN triggering mechanism.

    \item  The ratio of narrow-line to broad-line AGNs in clusters is comparable to that in the field.
    Interestingly, three out of four broad-line AGNs, including the only QSO, are found close to the cluster centre (within $r_{500}$). SED analysis of these broad-line AGNs revealed that their accretion power is among the highest in our sample. These results are possibly indicating AGN triggering in infalling galaxies by RPS.

\end{itemize}

Future research should focus on maximizing AGN sample sizes within meticulously characterised cluster datasets. Our future plans include thorough studies of AGNs in large X-ray-selected cluster samples, such as X-CLASS \citet{koulouridis21}, XCS \citep{Mehrtens2012,Giles2022} and eROSITA \citep{Bulbul2024}.

\begin{acknowledgements}
The authors extend their gratitude to the anonymous referee for the attentive review and valuable feedback. ED acknowledges financial support by the European Union’s Horizon 2020 programme “XMM2ATHENA” under grant agreement No 101004168. EK acknowledges support under the grant 5089 financed by IAASARS/NOA.  EP acknowledges financial support by the European Union's Horizon 2020 programme "XMM2ATHENA" under grant agreement No 101004168. The research leading to these results has received funding (EP) from the European Union's Horizon 2020 Programme under the AHEAD2020 project (grant agreement n. 871158). This research has made use of "Aladin sky atlas" developed at CDS, Strasbourg Observatory, France. This research has made use of ESASky \citep{Baines2017,Giordano2018}, developed by the ESAC Science Data Centre (ESDC) team and maintained alongside other ESA science mission's archives at ESA's European Space Astronomy Centre (ESAC, Madrid, Spain). This research made use of Astropy, a community-developed core Python package for Astronomy (http://www.astropy.org, \citet{Astropy2022}). This publication made use of SAOImageDS9 \citep{Joye2003}. This publication made use of TOPCAT \citep{taylor2005} for table manipulations. The plots in this publication were produced using Matplotlib, a Python library for publication quality graphics \citep{Hunter2007}. Based on observations
obtained with {\it XMM-Newton}, an ESA science mission with instruments and contributions directly funded by ESA member states and NASA.
\end{acknowledgements}


%
%
\bibliographystyle{aa}
\bibliography{XXLAGN}

\begin{appendix}
\section{Inventory of host galaxies of X-ray detected AGNs in clusters}
Table ~\ref{table:list} provides a list of the 82 X-ray detected AGNs in our cluster sample, while Table ~\ref{fig:mosaic1} display their colour images.
\onecolumn
\begin{longtable}{ccccccccc}
\caption{82 X-ray detected AGNs in clusters within 4$r_{500}$ radius.} \\
\label{table:list}
\endfirsthead
\caption{Continued.}\\
\hline
\\
\endhead
\hline
\endfoot
Cluster name & $r_{500}$ & source name &RA (J2000) & Dec (J2000) & spec/phot & $z$ & host & Spectral \\
XLSSC& & 4XLSSU & degrees & degrees &  & & morphology & type \\
\multicolumn{1}{c}{(1)}& \multicolumn{1}{c}{(2)}&\multicolumn{1}{c}{(3)}&\multicolumn{1}{c}{(4)}&\multicolumn{1}{c}{(5)}&\multicolumn{1}{c}{(6)}&\multicolumn{1}{c}{(7)}&\multicolumn{1}{c}{(8)}&\multicolumn{1}{c}{(9)}\\
\hline
\\
025	&1	& J022521.1-043950 &36.337835       & -4.663836    &   spec&    0.265 & E &NL \\
030 &1  & J022310.5-041249		&35.795024	& -4.214393	&   spec&    0.626	& M	& ALG \\
040	&1	& J022206.2-043251 &35.523711	& -4.546156	&   spec&	 0.323	&E	&ALG \\
 080	&1	&	J021819.1-052343 &34.579399	& -5.396029	&   spec&	 0.647	&E	&ALG \\
 082	&1	&	 J021046.1-060854 &32.692832	& -6.148447	&   spec&	 0.428	&E	&NL \\
 091	&1	&	J023147.2-045702&37.947556	& -4.950985	&   spec&	 0.190	&M	&NL \\
 111	&1	&	J021234.2-053545 &33.142287	& -5.595251	&   spec&	 0.302	&S	&NL \\
 114	&1	&	 J020139.1-050118&30.413549	& -5.021844	&   spec&	 0.231	&E	&BL \\
 117	&1	&	 J021235.9-053210&33.14998	& -5.536541	&   spec&	 0.299	&E	&BL \\
 142	&1	&	 J021856.0-052611&34.733459	& -5.435817	&   spec&	 0.448	&E	& U\\
 168	&1	&	 J022935.0-055210&37.396398	& -5.869275	&   spec&	 0.293	&U	&QSO BL \\
 183	&1	&	 J022016.9-045645&35.070248	& -4.946167	&   spec&	 0.517	&E	& NL \\
 187	&1	&	 J021631.8-042958&34.132286	& -4.499856	&   spec&	 0.454	&E	& ALG\\
 194	&1	&	 J021648.5-043318&34.202471	& -4.555327	&   spec&	 0.41	&S/tidal& NL \\
 040	&2  &	 J022206.5-042909&35.52714	& -4.485877	&   spec&	 0.315	&E	& ALG \\
 049	&2	&	 J022347.1-043346&35.946002	& -4.563383	&   spec&	 0.49	&D	& NL \\
 071	&2	&	 J022238.1-050100&35.659184	& -5.016476	&   phot &	 0.87	&E	&-\\
 071	&2	&	 J022236.1-050143&35.650138	& -5.027919	&   spec&	 0.844	&D	&NL \\
 083	&2	&	 J021035.8-061027&32.649452	& -6.17388	&   spec&	 0.433	&E	&NL \\
 089	&2	&  J022832.6-044608 &37.134301	& -4.768676	&   phot &	 0.60	&E	&-\\
 097	&2  &	 J021325.6-060329&33.357639	& -6.058701	&   spec&	 0.694	&U	&U\\
 101	&2  &	 J020838.3-042511&32.15988	& -4.419889	&   spec&	 0.753	&E	&NL \\
 105	&2  &	 J023324.9-053106&38.35469	& -5.517847	&   spec&	 0.436	&M	& NL \\
 107	&2  &	 J020534.3-073708&31.392291	& -7.618196	&   phot &	 0.43	&no HSC	& - \\
 110	&2  &	 J021413.9-053405&33.557274	& -5.568639	&   phot &	 0.440	&no HSC&NL \\
 116	&2  &	 J021047.8-060354&32.699967	& -6.064865	&   phot &	 0.53	&E	&-\\ 
 130	&2  &	 J022053.8-052538&35.224761	& -5.42698	&   phot &	 0.54	&E	&-\\
 156	&2  &	 J020305.0-070948&30.771721	& -7.163623	&   phot &	 0.33	&D/tidal	& -\\
 158	&2  &	 J021118.8-042516&32.828806	& -4.422185	&   spec&	 0.445	&S/D	&BL \\
 048	&3	& J022236.1-032639&35.651316	& -3.444693	&   spec&	 1.008	&E	&NL\\
 083	&3	& J021117.5-061916&32.825396	& -6.320064	&   phot &	 0.43	&E	&-\\
 083	&3	& J021052.9-061809&32.721253	& -6.302944	&   spec&	 0.423	&E	&NL \\
 085	&3	& J021125.7-061936&32.858153	& -6.327543	&   phot &	 0.42	&U	&-\\
 085	&3	& J021153.7-061033&32.974242	& -6.177046	&   spec&	 0.420	&E	& U\\
 101	&3	& J020853.8-042937&32.225226	& -4.493658	&   phot &	 0.75	&E	&-\\
 139	&3	& J021649.4-032831&34.206554	& -3.474726	&   phot &	 0.21	&E/tidal &-\\	
 157	&3	& J020336.4-070010&30.901106	& -7.003745	&   phot &	 0.59	&E	&-\\ 
 159	&3	& J020919.5-051152&32.329839	& -5.198636	&   spec&	 0.612	&M	& NL \\
 163	&3	& J021007.1-060459&32.529514	& -6.083835	&   spec&	 0.282	&E	&NL \\
 172	&3	& J020613.0-054957&31.554222	& -5.832453	&   phot &	 0.43	&E	&-\\
 183	&3	&  J022029.8-044657&35.123919	& -4.782901	&   phot &	 0.51	&E/D	&-\\
 200	&3	&  J020115.4-064331&30.314141	& -6.725744	&   phot &	 0.32	&E	&-\\
 001	&4	&  J022445.4-035509 &36.188901	& -3.919289	&   spec&	 0.605	&U	&QSO BL \\
 003	&4	&  J022750.3-032106 &36.958818	& -3.352099	&   phot &	 0.84	&U	&-\\
 008	&4	&  J022519.4-035444 &36.331233	& -3.912253	&   spec&	 0.299	&ring/D	& BL \\
 011	&4	&  J022700.7-042022 &36.753235	& -4.339044	&   spec&	 0.053	&E	&NL \\
 018	&4	&  J022430.5-050842 &36.127355	& -5.144935	&   spec&	 0.322	&S/tidal	&NL \\
 029	&4	&  J022418.9-041316 &36.078919	& -4.221791	&   spec&	 1.057	&E	&U\\
 030	&4	&  J022254.3-041629 &35.725574	& -4.274752	&   spec&	 0.63	&E	&NL \\
 056	&4	&  J021537.1-045005 &33.904892	& -4.834451	&   spec&	 0.350	&E	&ALG \\
 064	&4	&  J021818.4-045843 &34.576028	& -4.978517	&   phot &	 0.87	&E	&-\\
 067	&4	&  J021835.9-053758 &34.649637	& -5.632743	&   spec&	 0.387	&E	&BL\\
 071	&4	&  J022255.1-045328 &35.730888	& -4.891314	&   phot &	 0.83	&E/U	&-\\
 077	&4	& J021731.0-032444  &34.380138	& -3.41251	&   phot &	 0.20	&S	&-\\
 078	&4	&  J021616.8-045033 &34.069856	& -4.842637	&   spec&	 0.959	&E/M	&NL \\
 078	&4	&  J021610.6-045232 &34.044313	& -4.874921	&   spec&	 0.956	&E/D	&NL \\
 091	&4	&  J023138.0-051420 &37.908463	& -5.238549	&   spec&	 0.187	&E	&NL \\
 093	&4	&  J020629.7-064905 &31.624312	& -6.81841	&   phot &	 0.42	&E	&-\\
 093	&4	& J020614.4-065635  &31.561428	& -6.94357	&   phot &	 0.43	&U	&-\\
 097	&4	&  J021304.4-060037 &33.268811	& -6.009716	&   phot &	 0.69	&E	&-\\
 104	&4	& J022841.1-055724  &37.171036	& -5.956224	&   spec&	 0.297	& E &ALG\\
 106	&4	& J020516.2-055230  &31.317255	& -5.875662	&   spec&	 0.295	&E	&NL\\
 107	&4	&  J020553.3-073743 &31.473609	& -7.627636	&   phot &	 0.43	&no HSC&\\
 109	&4	& J020909.6-062617 &32.289923	& -6.437786	&   phot &	 0.49	&E	&-\\
 109	&4	&  J020930.7-062542  &32.378151	& -6.428463	&   phot &	 0.49	&E	&- \\
 110	&4	&  J021353.0-053917 &33.470966	& -5.655553	&   phot &	 0.45	&E	&-\\
 116	&4	&  J020954.7-060712 &32.479197	& -6.120809	&   phot &	 0.53	&U	&-\\
 124	&4	&  J021759.0-045551 &34.496463	& -4.931106	&   spec&	 0.506	&E/tidal & ALG\\	
 128	&4	&  J022342.4-030411 &35.927244	& -3.069659	&   phot &	 0.48	&E	&-\\
 135	&4	&  J021453.6-035929 &33.724019	& -3.991403	&   phot &	 0.37	&U	&-\\
 137	&4	&  J021815.7-034141 &34.565714	& -3.694404	&   phot &	 0.29	&U &-\\
 149	&4	&  J022959.9-045716 &37.498909	& -4.954494	&   spec&	 0.286	&U & NL \\
 158	&4	&  J021144.9-041843 &32.937205	& -4.311633	&   phot &	 0.44	&E/D	&-\\
 158	&4	&  J021043.2-042509 &32.679161	& -4.420212	&   phot &	 0.45	&E	&-\\
 169	&4	&  J023044.0-053605 &37.68341	& -5.601003	&   phot &	 0.49	&S &-\\
 169	&4	&  J023007.4-054902 &37.530904	& -5.81804	&   phot &	 0.49	&E &-\\
 182	&4	&  J022542.5-032024 &36.426675	& -3.340322	&   spec&	 0.170	&E &BL\\
 183	&4	&  J021922.3-045943 &34.844415	& -4.995945	&   phot &	 0.51	&E	&-\\
 183	&4	&  J021921.5-045838 &34.840431	& -4.977854	&   spec&	 0.512	&E	&U\\
 184	&4	& J022115.7-040901  &35.315354	& -4.150097	&   phot &	 0.81	&U	&-\\
 187	&4	&  J021543.7-042456 &33.932907	& -4.414839	&   spec&	 0.457	&E	& ALG\\
 199	&4	&  J020019.5-064750 &30.082133	& -6.797771	&   phot &	 0.33	&  E    &-\\		
											
\end{longtable}
\tablefoot{(1) XXL Cluster name, (2) $r_{500}$ annulus where the X-ray detected AGNs is located, (3) X-ray point-source ID (4) right ascension of the optical counterpart, (5) declination of the optical counterpart, (6) flag for spectroscopic (spec) or photometric (phot) redshift, (7) redshift, (8) morphological classification of the host galaxy, (9) optical spectral classification: narrow-line (NL), broad-line(BL), absorption-line galaxy (ALG), undefined (U)}

\begin{longtable}{cccccc}
\caption{RGB colour images of X-ray-detected AGN host galaxies, based on the gri filters of HSC.} \\
\label{table:images}
\endfirsthead
\caption{Continued.}\\
\hline
\\
\endhead
\hline
\\
J022521&J022310&J022206&J021819&J021046&J023147\\
    \vspace{0.3cm}
    \includegraphics[trim={0.5cm 0.5cm 0.5cm 0.5cm},clip,width=0.14\textwidth,height=0.14\textwidth]
    {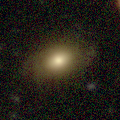}&\includegraphics[trim={0.8cm 0.8cm 0.8cm 0.8cm},clip,width=0.14\textwidth,height=0.14\textwidth]{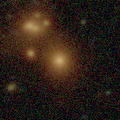}&\includegraphics[trim={0.8cm 0.8cm 0.8cm 0.8cm},clip,width=0.14\textwidth,height=0.14\textwidth]{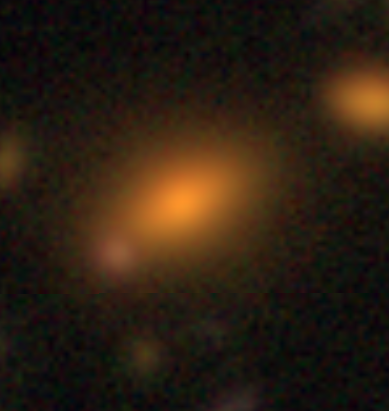}&\includegraphics[trim={1.2cm 1.2cm 1.2cm 1.2cm},clip,width=0.14\textwidth,height=0.14\textwidth]{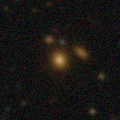}&\includegraphics[trim={0.7cm 0.7cm 0.7cm 0.7cm},clip,width=0.14\textwidth,height=0.14\textwidth]{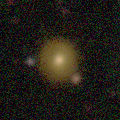}& \includegraphics[trim={0.7cm 0.7cm 0.7cm 0.7cm},clip,width=0.14\textwidth,height=0.14\textwidth]{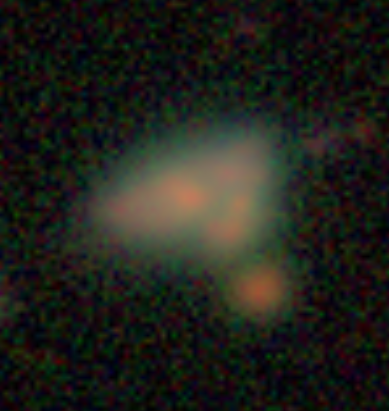}\\    
    J021234&J020139&J021235&J021856&J022935&J022016\\
        \vspace{0.3cm}
    \includegraphics[trim={0.4cm 0.4cm 0.4cm 0.4cm},clip,width=0.14\textwidth,height=0.14\textwidth]{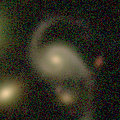}&
    \includegraphics[trim={0.8cm 0.8cm 0.8cm 0.8cm},clip,width=0.14\textwidth,height=0.14\textwidth]{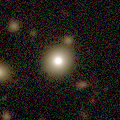}&
    \includegraphics[trim={0.5cm 0.5cm 0.5cm 0.5cm},clip,width=0.14\textwidth,height=0.14\textwidth]{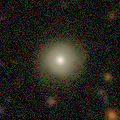}&\includegraphics[trim={1.5cm 1.5cm 1.5cm 1.5cm},clip,width=0.14\textwidth,height=0.14\textwidth]{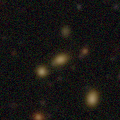}&\includegraphics[trim={0.5cm 0.5cm 0.5cm 0.5cm},clip,width=0.14\textwidth,height=0.14\textwidth]{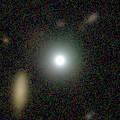}&
    \includegraphics[trim={1.1cm 1.1cm 1.1cm 1.1cm},clip,width=0.14\textwidth,height=0.14\textwidth]{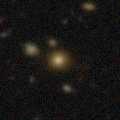}\\
    J021631&J021648&J022206&J022347&J022238&J022236\\
            \vspace{0.3cm}
    \includegraphics[trim={1cm 1cm 1cm 1cm},clip,width=0.14\textwidth,height=0.14\textwidth]{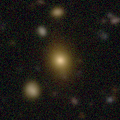}
    &\includegraphics[trim={0.5cm 0.5cm 0.5cm 0.5cm},clip,width=0.14\textwidth,height=0.14\textwidth]{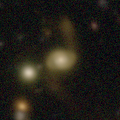}&
    \includegraphics[trim={0.7cm 0.7cm 0.7cm 0.7cm},clip,width=0.14\textwidth,height=0.14\textwidth]{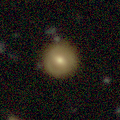}&\includegraphics[trim={0.8cm 0.8cm 0.8cm 0.8cm},clip,width=0.14\textwidth,height=0.14\textwidth]{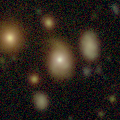}&\includegraphics[trim={1.2cm 1.2cm 1.2cm 1.2cm},clip,width=0.14\textwidth,height=0.14\textwidth]{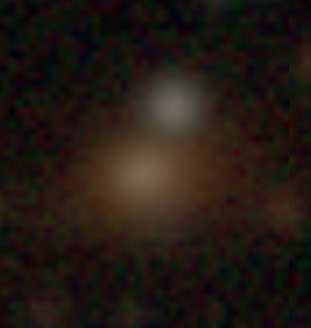}&
    \includegraphics[trim={1.2cm 1.2cm 1.2cm 1.2cm},clip,width=0.14\textwidth,height=0.14\textwidth]{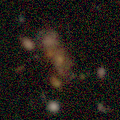}\\
            J021035&J022832&J021325&020838&J023324&J021047\\
              \vspace{0.3cm}
    \includegraphics[trim={1cm 2cm 2cm 2cm},clip,width=0.14\textwidth,height=0.14\textwidth]{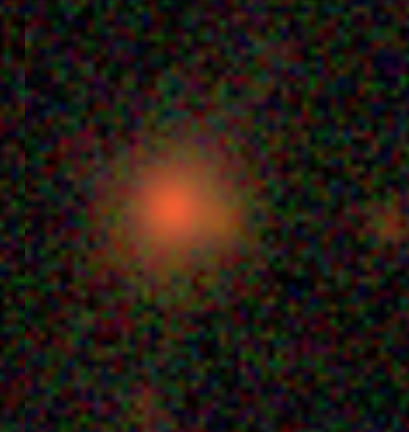}&\includegraphics[trim={1.5cm 1.5cm 1.5cm 1.5cm},clip,width=0.14\textwidth,height=0.14\textwidth]{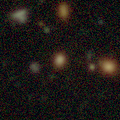}&\includegraphics[trim={0.2cm 0.2cm 0.2cm 0.2cm},clip,width=0.14\textwidth,height=0.14\textwidth]{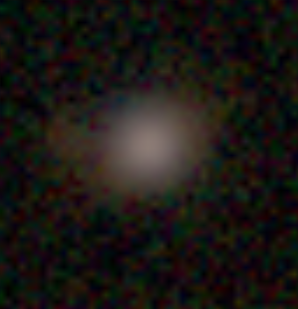}&
    \includegraphics[trim={1.5cm 1.5cm 1.5cm 1.5cm},clip,width=0.14\textwidth,height=0.14\textwidth]{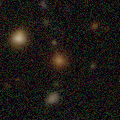}&\includegraphics[trim={1.2cm 1.2cm 1.2cm 1.2cm},clip,width=0.14\textwidth,height=0.14\textwidth]{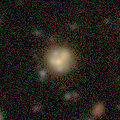}&\includegraphics[trim={1.5cm 1.5cm 1.5cm 1.5cm},clip,width=0.14\textwidth,height=0.14\textwidth]{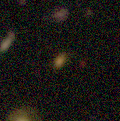}\\
            J022053&J020305&J021118&J022236&J021117&J021052\\
            \vspace{0.3cm}
    \includegraphics[trim={1.5cm 1.5cm 1.5cm 1.5cm},clip,width=0.14\textwidth,height=0.14\textwidth]{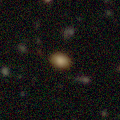}&\includegraphics[trim={0.6cm 0.6cm 0.6cm 0.6cm},clip,width=0.14\textwidth,height=0.14\textwidth]{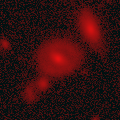}&\includegraphics[trim={0.7cm 0.7cm 0.7cm 0.7cm},clip,width=0.14\textwidth,height=0.14\textwidth]{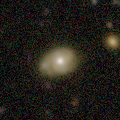}&
    \includegraphics[trim={0.1cm 0.1cm 0.1cm 0.1cm},clip,width=0.14\textwidth,height=0.14\textwidth]{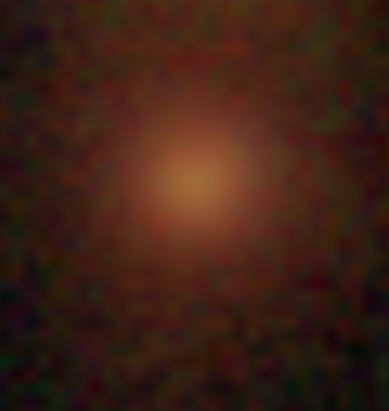}&
    \includegraphics[trim={1.5cm 1.5cm 1.5cm 1.5cm},clip,width=0.14\textwidth,height=0.14\textwidth]{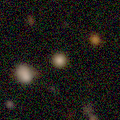}&\includegraphics[trim={1.5cm 1.5cm 1.5cm 1.5cm},clip,width=0.14\textwidth,height=0.14\textwidth]{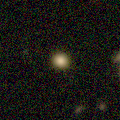}\\
                J021125&J021153&J020853&J021649&J020336&J020919\\
            \vspace{0.3cm}
     \includegraphics[trim={1.6cm 1.6cm 1.6cm 1.6cm},clip,width=0.14\textwidth,height=0.14\textwidth]{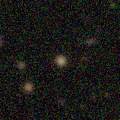}&\includegraphics[trim={0.9cm 0.9cm 0.9cm 0.9cm},clip,width=0.14\textwidth,height=0.14\textwidth]{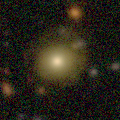}&\includegraphics[trim={1.7cm 1.7cm 1.7cm 1.7cm},clip,width=0.14\textwidth,height=0.14\textwidth]{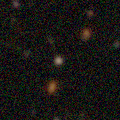}&\includegraphics[trim={0.2cm 0.2cm 0.2cm 0.2cm},clip,width=0.14\textwidth,height=0.14\textwidth]{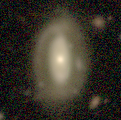}&\includegraphics[trim={1.5cm 1.5cm 1.5cm 1.5cm},clip,width=0.14\textwidth,height=0.14\textwidth]{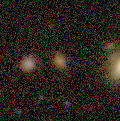}&\includegraphics[trim={1.2cm 1.2cm 1.2cm 1.2cm},clip,width=0.14\textwidth,height=0.14\textwidth]{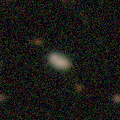}\\
                    J021007&J020613&J022029&J020115&J022445&J022750\\
            \vspace{0.3cm}
     \includegraphics[trim={1.1cm 1.1cm 1.1cm 1.1cm},clip,width=0.14\textwidth,height=0.14\textwidth]{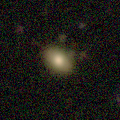}&
    \includegraphics[trim={1.5cm 1.5cm 1.5cm 1.5cm},clip,width=0.14\textwidth,height=0.14\textwidth]{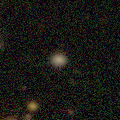}&\includegraphics[trim={1.4cm 1.4cm 1.4cm 1.4cm},clip,width=0.14\textwidth,height=0.14\textwidth]{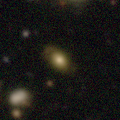}&\includegraphics[trim={1.1cm 1.1cm 1.1cm 1.1cm},clip,width=0.14\textwidth,height=0.14\textwidth]{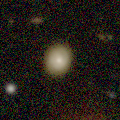}&\includegraphics[trim={1cm 1cm 1cm 1cm},clip,width=0.14\textwidth,height=0.14\textwidth]{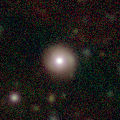}&\includegraphics[trim={1.6cm 1.6cm 1.6cm 1.6cm},clip,width=0.14\textwidth,height=0.14\textwidth]{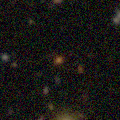}\\
                    J022519&J2022700&J022430&J022418&J022254&J021537\\
            \vspace{0.3cm}
    \includegraphics[trim={0.9cm 0.9cm 0.9cm 0.9cm},clip,width=0.14\textwidth,height=0.14\textwidth]{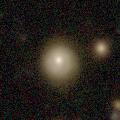}&
    \includegraphics[trim={0.4cm 0.4cm 0.4cm 0.4cm},clip,width=0.14\textwidth,height=0.14\textwidth]{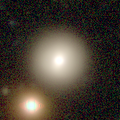}&\includegraphics[trim={0.6cm 0.6cm 0.6cm 0.6cm},clip,width=0.14\textwidth,height=0.14\textwidth]{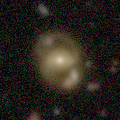}&\includegraphics[trim={1.6cm 1.6cm 1.6cm 1.6cm},clip,width=0.14\textwidth,height=0.14\textwidth]{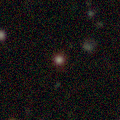}&\includegraphics[trim={1.5cm 1.5cm 1.5cm 1.5cm},clip,width=0.14\textwidth,height=0.14\textwidth]{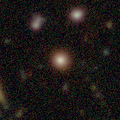}&\includegraphics[trim={0.5cm 0.5cm 0.5cm 0.5cm},clip,width=0.14\textwidth,height=0.14\textwidth]{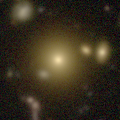}\\
                    J021818&J021835&J022255&J021731&J021616&J021610\\
            \vspace{0.3cm}
    \includegraphics[trim={1.5cm 1.5cm 1.5cm 1.5cm},clip,width=0.14\textwidth,height=0.14\textwidth]{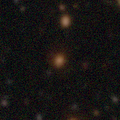}&\includegraphics[trim={1.1cm 1.1cm 1.1cm 1.1cm},clip,width=0.14\textwidth,height=0.14\textwidth]{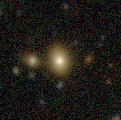}&\includegraphics[trim={1.5cm 1.5cm 1.5cm 1.5cm},clip,width=0.14\textwidth,height=0.14\textwidth]{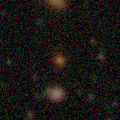}&\includegraphics[trim={0.4cm 0.4cm 0.4cm 0.4cm},clip,width=0.14\textwidth,height=0.14\textwidth]{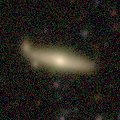}&\includegraphics[trim={1.3cm 1.3cm 1.3cm 1.3cm},clip,width=0.14\textwidth,height=0.14\textwidth]{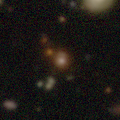}&\includegraphics[trim={1.1cm 1.1cm 1.1cm 1.1cm},clip,width=0.14\textwidth,height=0.14\textwidth]{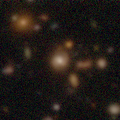}\\
                        J023138&J020629&J020614&J021304&J022841&J020516\\
            \vspace{0.3cm}
    \includegraphics[trim={1cm 1cm 1cm 1cm},clip,width=0.14\textwidth,height=0.14\textwidth]{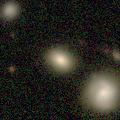}&\includegraphics[trim={1.5cm 1.5cm 1.5cm 1.5cm},clip,width=0.14\textwidth,height=0.14\textwidth]{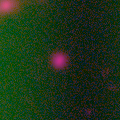}&\includegraphics[trim={1.5cm 1.5cm 1.5cm 1.5cm},clip,width=0.14\textwidth,height=0.14\textwidth]{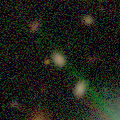}&\includegraphics[trim={1.5cm 1.5cm 1.5cm 1.5cm},clip,width=0.14\textwidth,height=0.14\textwidth]{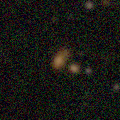}&\includegraphics[trim={0.5cm 0.5cm 0.5cm 0.5cm},clip,width=0.14\textwidth,height=0.14\textwidth]{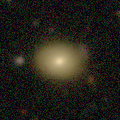}&\includegraphics[trim={1.2cm 1.2cm 1.2cm 1.2cm},clip,width=0.14\textwidth,height=0.14\textwidth]{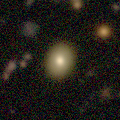}\\
                        J020909&J020930&J021353&J020954&J021759&J022342\\
            \vspace{0.3cm}
    \includegraphics[trim={1.5cm 1.5cm 1.5cm 1.5cm},clip,width=0.14\textwidth,height=0.14\textwidth]{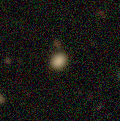}&\includegraphics[trim={1.3cm 1.3cm 1.3cm 1.3cm},clip,width=0.14\textwidth,height=0.14\textwidth]{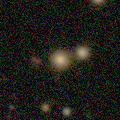}&\includegraphics[trim={1.5cm 1.5cm 1.5cm 1.5cm},clip,width=0.14\textwidth,height=0.14\textwidth]{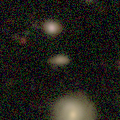}&\includegraphics[trim={1.5cm 1.5cm 1.5cm 1.5cm},clip,width=0.14\textwidth,height=0.14\textwidth]{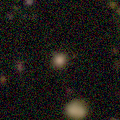}&\includegraphics[trim={0.6cm 0.6cm 0.6cm 0.6cm},clip,width=0.14\textwidth,height=0.14\textwidth]{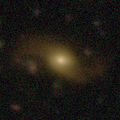}&\includegraphics[trim={1.5cm 1.5cm 1.5cm 1.5cm},clip,width=0.14\textwidth,height=0.14\textwidth]{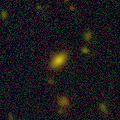}\\
                        J021453&J021815&J022959&J021144&J021043&J023044\\
            \vspace{0.3cm}
    \includegraphics[trim={1.5cm 1.5cm 1.5cm 1.5cm},clip,width=0.14\textwidth,height=0.14\textwidth]{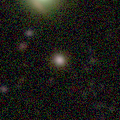}&\includegraphics[trim={3.5cm 2.5cm 2.5cm 4.5cm},clip,width=0.14\textwidth,height=0.14\textwidth]{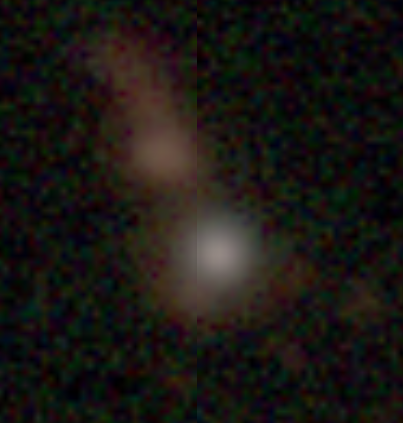}&\includegraphics[trim={1cm 1cm 1cm 1cm},clip,width=0.14\textwidth,height=0.14\textwidth]{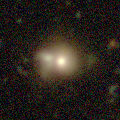}&\includegraphics[trim={0.4cm 0.4cm 0.4cm 0.4cm},clip,width=0.14\textwidth,height=0.14\textwidth]{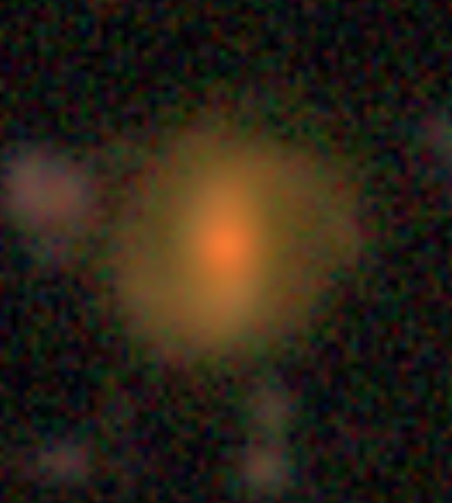}&\includegraphics[trim={1.5cm 1.5cm 1.5cm 1.5cm},clip,width=0.14\textwidth,height=0.14\textwidth]{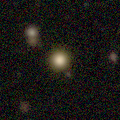}&\includegraphics[trim={0.6cm 0.6cm 0.6cm 0.6cm},clip,width=0.14\textwidth,height=0.14\textwidth]{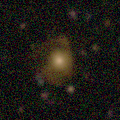}\\
                            J023007&J022542&J021922&J021921&J022115&J021543\\
            \vspace{0.3cm}
    \includegraphics[trim={1.3cm 1.3cm 1.3cm 1.3cm},clip,width=0.14\textwidth,height=0.14\textwidth]{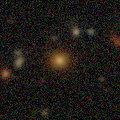}&\includegraphics[trim={1.5cm 1.5cm 1.5cm 1.5cm},clip,width=0.14\textwidth,height=0.14\textwidth]{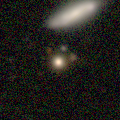}&\includegraphics[trim={1.4cm 1.4cm 1.4cm 1.4cm},clip,width=0.14\textwidth,height=0.14\textwidth]{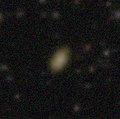}&\includegraphics[trim={1.5cm 1.5cm 1.5cm 1.5cm},clip,width=0.14\textwidth,height=0.14\textwidth]{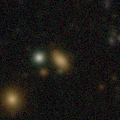}&\includegraphics[trim={1.7cm 1.7cm 1.7cm 1.7cm},clip,width=0.14\textwidth,height=0.14\textwidth]{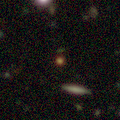}&\includegraphics[trim={1cm 1cm 1cm 1cm},clip,width=0.14\textwidth,height=0.14\textwidth]{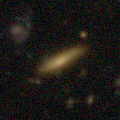}\\
    J020019.&&&&&\\
            \vspace{0.3cm}
    \includegraphics[trim={1.2cm 1.2cm 1.2cm 1.2cm},clip,width=0.14\textwidth,height=0.14\textwidth]{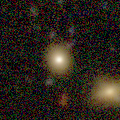}\\
    
    \label{fig:mosaic1}
\end{longtable}

\twocolumn

\section{The case of spiral galaxies}
\label{app:spiral}

\begin{figure}[htp]
    \centering
    \includegraphics[width=8.2cm]{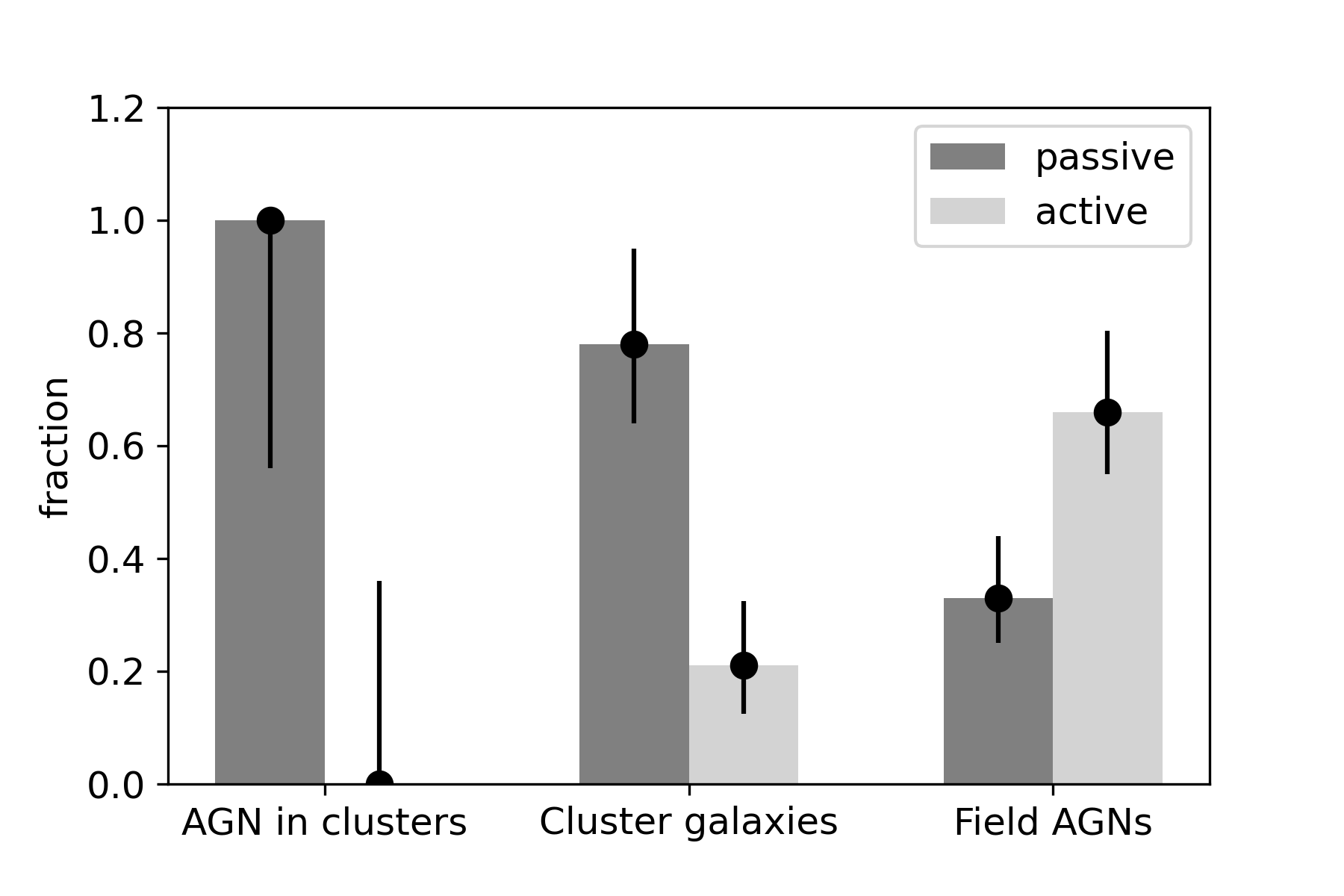}
    \caption{Fraction of passive vs active spiral host galaxies in the main sample of X-ray AGNs in clusters comparing with both the control samples, with cluster galaxies and with X-ray AGNs from the field. The classification has been made based on visual inspection from the HSC images. The sample
    is statistically small but there is a clear trend, specifically in the case of the X-ray AGNs in clusters we find only passive spiral AGN host galaxies and found up to 2r500 cluster region. Error bars indicate the $ 1 \sigma $ confidence limits for small numbers of events
    \citep{Gehrels1986}.}
    \label{fig:spirals}
\end{figure}

In this section, we focus specifically on the spiral host galaxies of X-ray AGNs in clusters, comparing them with the spiral galaxy population in the control samples. We categorised them into two types: passive (red) and active (blue). The colours were defined using HSC $gri$ photometric magnitudes. We define a galaxy as red and thus passive when $g-r$ > 0.5 \citep{daoutis, shimaka}. Passive spiral galaxies are characterised by a lack of star formation, in sharp contrast to active spiral galaxies \citep{shimaka}, which are also the vast majority. Our results are presented in Fig.~\ref{fig:spirals}. Despite the limited number of spiral host cases, we observe a clear trend in the results. Specifically, in the main sample of X-ray AGNs in clusters, all spiral hosts are passive, whereas in the field, active hosts dominate. We also examined the spiral host galaxies in the control sample of cluster galaxies, finding mostly passive cases with only a few active ones. Additionally, we reviewed the SFR of our sample galaxies using their SED, as described in Sect.~\ref{sec:sed}. Because of the low number of spiral hosts, we use all derived SEDs even if $\rm \chi^2_r>5$. The results confirm the low-level of star forming activity (average SFR $\approx 0.8 M_\odot$/yr) in the spiral galaxies of our sample. 

This is likely a consequence of environmental processes such as ram pressure stripping. Quenching mechanisms probably suppress star formation more rapidly than the time needed for morphological transformation through bulge growth and disc fading \citep[e.g.][]{Kelkar2019,Martinez2022,Oxland2024}. These results suggest that the conditions within clusters, including interactions with other galaxies and the intracluster medium, play a significant role in quenching star formation and rendering these spiral hosts inactive. 

We conclude that all spiral X-ray AGN hosts within clusters are passive in terms of star formation, as defined by colour or SED analysis. This is also true for the majority of non-AGN spiral cluster galaxies, but opposite from what we observe in the field. Although the number of spiral hosts is small, we argue that this is likely a result of ram pressure stripping.

\begin{figure}[htp]
    \centering
    \includegraphics[width=5cm]{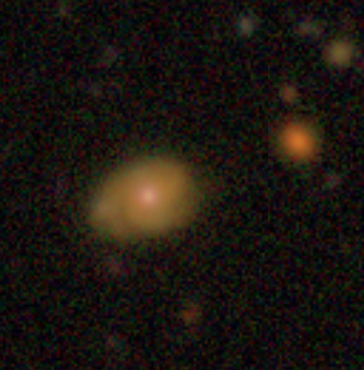} \includegraphics[width=5cm]{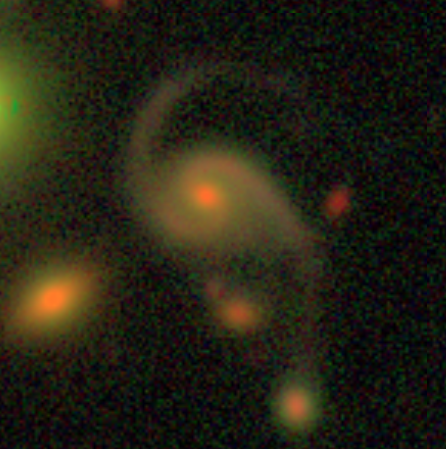} \includegraphics[width=5cm]{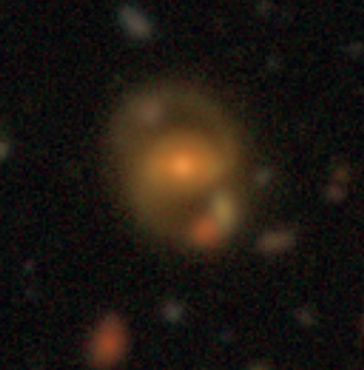}

   \caption{Passive spiral host galaxies of X-ray AGNs. RGB colour images are produced by \textit{gri} broad-band imaging from the HSC-SSP PDR3. From top to bottom: spectroscopically confirmed cluster members at redshift $z$=0.445 (4XLSSU J021118.8-042516), $z$=0.302 (4XLSSU J021234.2-053545) and $z$=0.322 (4XLSSU J022430.5-050842). The image scale is approximately 30"×30".}
    \label{fig:spirals2}
\end{figure}

\section{SFR-stellar mass relation of AGNs in clusters}
\label{app:SFR}

In this section we explore the SFR-stellar mass relation of our sources relative to the main sequence (MS) of the star-forming galaxies. Star-forming galaxies show a tight correlation between their stellar mass (M$_*$) and SFRs, known as the MS of star-forming galaxies \citep[e.g., ][]{whitaker2012, speagle2014,schreiber2015}. This relation is valid through a wide range of redshifts \citep{schreiber2016}. Hence, we used the SED derived properties of our samples to examine the location of the X-ray AGNs with respect to the MS.

\begin{figure}
       \includegraphics[width=0.48\textwidth]{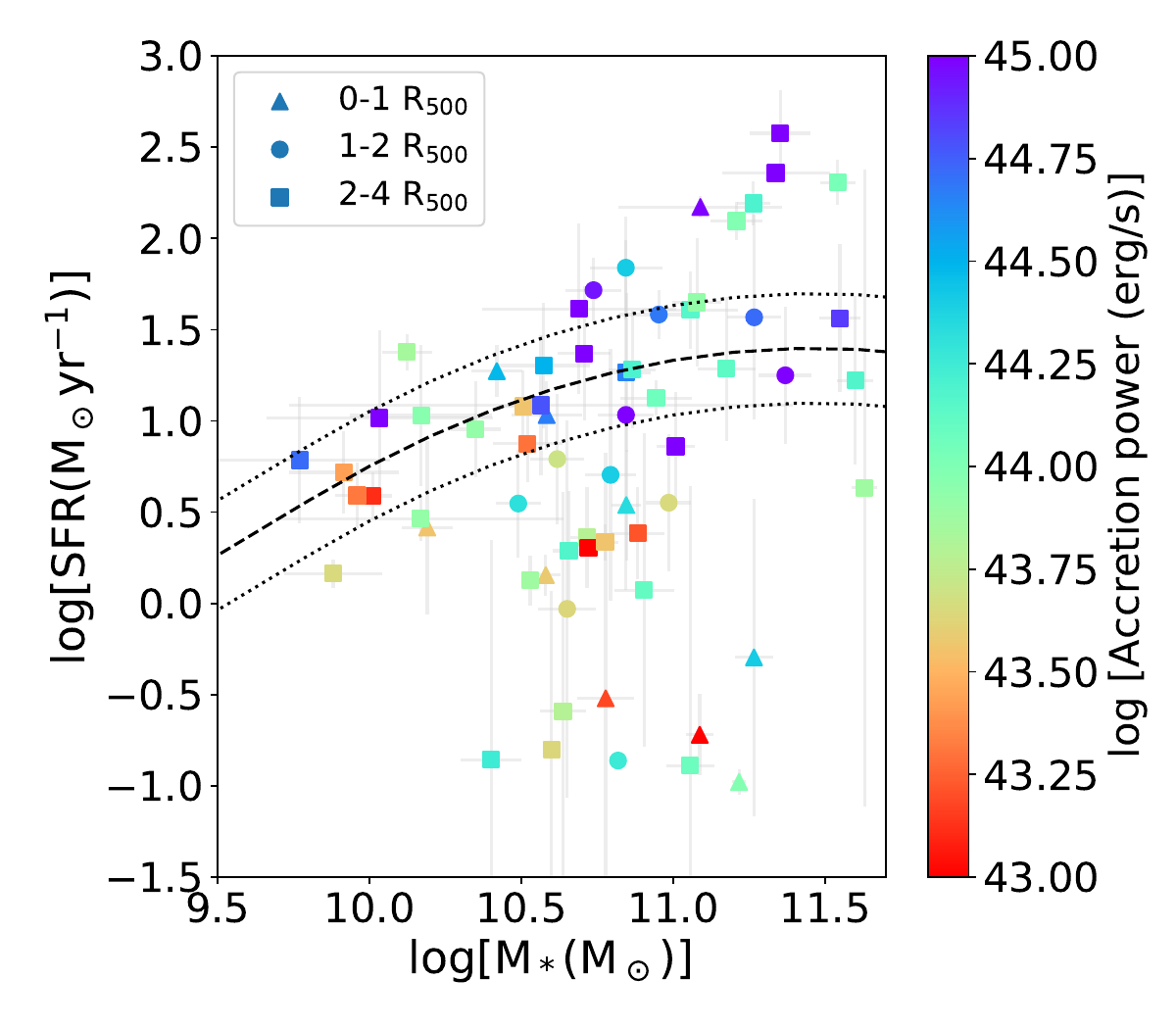} 
\caption{SFR as a function of M$_*$, colour-coded based on the AGN accretion power. The different shapes represent the AGN samples used in our analysis in different distances from the cluster centres as indicated in the legend. The dashed line represents the main sequence of star-forming galaxies obtained by \citet{schreiber2015} with median redshift value z=0.5. The dotted lines correspond to the uncertainties defined as $\pm0.3$ dex.}\label{m_SFR}
\end{figure}

In Fig.~\ref{m_SFR}, we plot our sources in the SFR-M$_*$ plane using the estimated SFR and M$_*$ values for our three samples, as indicated in the legend. The sources are colour-coded with the accretion power (AP) of the AGNs ("AGN.accretion.power" derived from CIGALE). It provides the intrinsic (unextinct) luminosity of the AGN disc averaged in all directions \citep[e.g.][]{yang2023}. We compared the source positions relative to the MS, using for the latter the analytical expression of equation 9 of \citet{schreiber2015} (dashed line). For this calculation, we used the median redshift of our sample ($\rm z_{med}=0.5$). The sources that have SFRs within $\rm 0.3$\,dex from the \citet{schreiber2015} SFR (dotted lines) are considered to lie within the MS. Using the best SFR and M$_*$ values, we find that the majority of the sources inside or above the MS (that is about 56\% of the full sample) have in general higher AGN accretion power $(2.8\times10^{44}$ erg/s) compared to those that lie below the MS $(6.8\times10^{43}$ erg/s). To evaluate whether the sources below and those above or inside the MS come from the same parent distribution, we performed a two-side Kolmogorov-Smirnov (KS) test. The p-value is 0.00053, indicating that the two distributions are different. These results agree with previous studies \citep[e.g.][]{aird2019,Pouliasis2022b} that may indicate that higher accreting SMBHs may reside in host galaxies with enhanced SFR.

In clusters this might be linked to the high frequency of merging and disturbed galaxies. Indeed, we found evidence that host galaxies that are visually identified as disturbed exhibit relatively high values of SFR. This elevated SFR appears to be associated with substantial AP. In more detail, the same hosts that show signs of disturbance, possibly due to interactions or mergers, are also the ones where both vigorous star formation and intense accretion activity are present. This is in agreement with previous results \citep[e.g.][]{Storchi2019,Silva2021,Cezar2024} and suggests a connection between the dynamic state of these galaxies and their energetic processes, both in forming new stars and in feeding their central black holes.

Furthermore, our results revealed that 50\% of the sources in the centre of clusters lie inside or above the MS of the star-forming galaxies, while this percentage increases to about 65\% in the case of sources in the field.

\end{appendix}
\end{document}